\documentclass{aa}  

\usepackage{graphicx}
\usepackage{txfonts}
\usepackage{amsfonts}
\usepackage{color}
\usepackage[dvipsnames]{xcolor}

\usepackage{hyperref}
\newcommand{\dd}{\mathrm{d}}

\newcommand{\hMsun}{h^{-1}\mathrm{M}_\odot}

\usepackage[normalem]{ulem}

\makeatletter
\renewcommand*\aa@pageof{, page \thepage{} of \pageref*{LastPage}}
\makeatother
\hypersetup{colorlinks=true,allcolors=[rgb]{0,0,0.8}}

\begin{document}

   \title{STRAWBERRY: Finding haloes in the gravitational potential}

   \subtitle{}

   \author{Tamara R. G. Richardson
          \inst{1}
          \and
          Jens St\"ucker\inst{2}
          \and
          Raul E. Angulo\inst{1,3}}

   \institute{Donostia International Physics Center (DIPC), Paseo Manuel de Lardizabal, 4, 20018, Donostia-San Sebasti\'an,
Gipuzkoa, Spain\\
             \email{tamara.richardson@dipc.org}
        \and
        Institute for Astronomy, University of Vienna, Türkenschanzstraße 17, Vienna 1180, Austria
        \and 
        IKERBASQUE, Basque Foundation for Science, E-48013, Bilbao, Spain
        }

   \date{Received MM DD, YYYY; accepted MM DD, YYYY}
 
  \abstract
  { 
Here, we present a novel algorithm that discriminates between bound and unbound particles by consideration of the gravitational potential from an accelerated reference frame -- also referred to as `the boosted potential'. Particles are considered bound if their energy does not exceed the escape energy of a potential well -- given by the closest saddle-point that connects to a deeper potential minimum. This approach has core benefits over previous approaches, since it does not require any ad-hoc thresholds (such as over-density criteria), it includes the gravitational effect of all particles in the binding criterion (improving over widely used self-potential binding checks) and it only operates with instantaneous information (making it simpler than approaches based on dynamical histories).
We show that particles typically become bound between their first peri- and apo-centeric passage and that bound and unbound populations show very distinct characteristics through their distribution in phase space, their density profiles, their virial ratios, and their redshift evolution. Our findings suggest that it is possible to understand haloes as two-component systems, with one component being bound, virialized, of finite extent and evolving slowly in quasi-equilibrium and the other component being unbound, unvirialized and evolving rapidly.

  }
  

   \maketitle

\section{Introduction}
\label{sec:intro}

In the standard $\Lambda$CDM cosmological model, dark matter haloes are one of the building blocks of non-linear structure. As such, a lot of research is built upon our understanding of these structures. A non-exhaustive list of examples includes, the halo model of the non-linear power spectrum and its extensions \citep[e.g.][]{Seljak2000,Mead2020,Asgari2023,Aycoberry2024, Salazar2024} which rely on knowing, how abundant haloes are and how matter is distributed within them; galaxy clustering studies \citep[e.g.][]{eBOSS_2021,DES_Y1,DES_Y3, DESI2024, Euclid2024} which are heavily reliant on the definition and properties of haloes in order to model the clustering properties of galaxies \citep[e.g.][]{Jing1998,Benson2000,Peacock2000,Vale2006,Conroy2006,Contreras2021,Ortega2024}; galaxy cluster number count analyses which are heavily dependent on the calibration of scaling relations between observable properties and halo properties \citep[e.g.][]{Planck_cluster_2016,Kids2021, Lesci2022,SPTpol2024,eRASS2024,CHEXMATE2025}; and dedicated `zoom-in' simulation runs of galaxy and structure formation \citep[e.g.][]{Hahn2017,Cui2018,Cui2022,wang2020_prof,Nadler2023,Pellisier2023,Nelson2024} which require a robust understanding of how haloes form to accurately select the Lagrangian patch that is re-simulated. Therefore it is evident that it is important to have a clear definition of haloes and their boundaries.

The simplest, and arguably the most widely used  boundary defines haloes as spherical regions which are denser than the cosmological background.
This results in a set of characteristic mass scales known as spherical overdensity masses, typically denoted $M_{\rm \Delta c}$ or $M_{\rm \Delta b}$, which represent the total mass contained within a spherical region enclosing an average density that is $\Delta$ times the critical or background matter density of the Universe. This definition is motivated by the theoretical description of virialised haloes within the spherical collapse framework. 
This definition has many benefits which have made it popular. First, it is intuitive and easy to implement. Second, it is flexible, with the possibility of varying the value of $\Delta$ depending on the application. For example, studies that focus on the innermost regions of haloes, such as the study of hot gas in the intra cluster medium \citep[e.g.][]{Planck_cluster_2016,ACT_clusters2021,SPTpol2024,eRASS2024,CHEXMATE2025}, will tend to use mass definitions corresponding to these regions, e.g. $M_{500 \rm c}$, $M_{2500 \rm c}$, while studies interested in larger radii, such as weak lensing mass estimates \citep[e.g][]{Sereno2015,Bellagamba2019,HSC_clusters2020,Lesci2022}, will tend to use masses defined at lower overdensities, e.g. $M_{200 \rm c}$ or $M_{200 \rm b}$. 

Nonetheless, this halo definition has several drawbacks. Notably, 
it enforces spherical symmetry while haloes are typically ellipsoidal, and present many small-scale features such as subhaloes or caustics \citep[e.g.][]{Eisenstein1995, Springel2004,Vera-Ciro2011, Despali2014,Bonnet2022}.
Moreover, spherical overdensities are not well suited to studying the evolution of haloes, due to the fact that the definition itself changes depending on the cosmological background, through an effect known as pseudo-evolution \citep{Diemer2013}. For example, this effect makes it more difficult to disentangle how much matter has been accreted over a given time period, as the spherical overdensity mass will increase over time even in the absence of accretion, simply because the background density is decreasing. 
This definition is also ill-suited to describe the abundance of haloes, as  expressed through the halo mass function (HMF). Indeed, excursion-set theory predicts that the HMF should have no explicit cosmology dependence when expressed in the correct units \citep{Press1974,Bond1991}, in other words, that the HMF should be universal. 
In practice, simulations have revealed this to not be the case \citep[e.g.][]{Tinker2008,Despali2016, Castro2021, Ondaro2022}. However, an open question remains, what is the origin of observed non universality? Is it generated through a physical process? or does it simply emerge from the mass definition?

\begin{figure}
    \centering
    \includegraphics[width=0.9\linewidth]{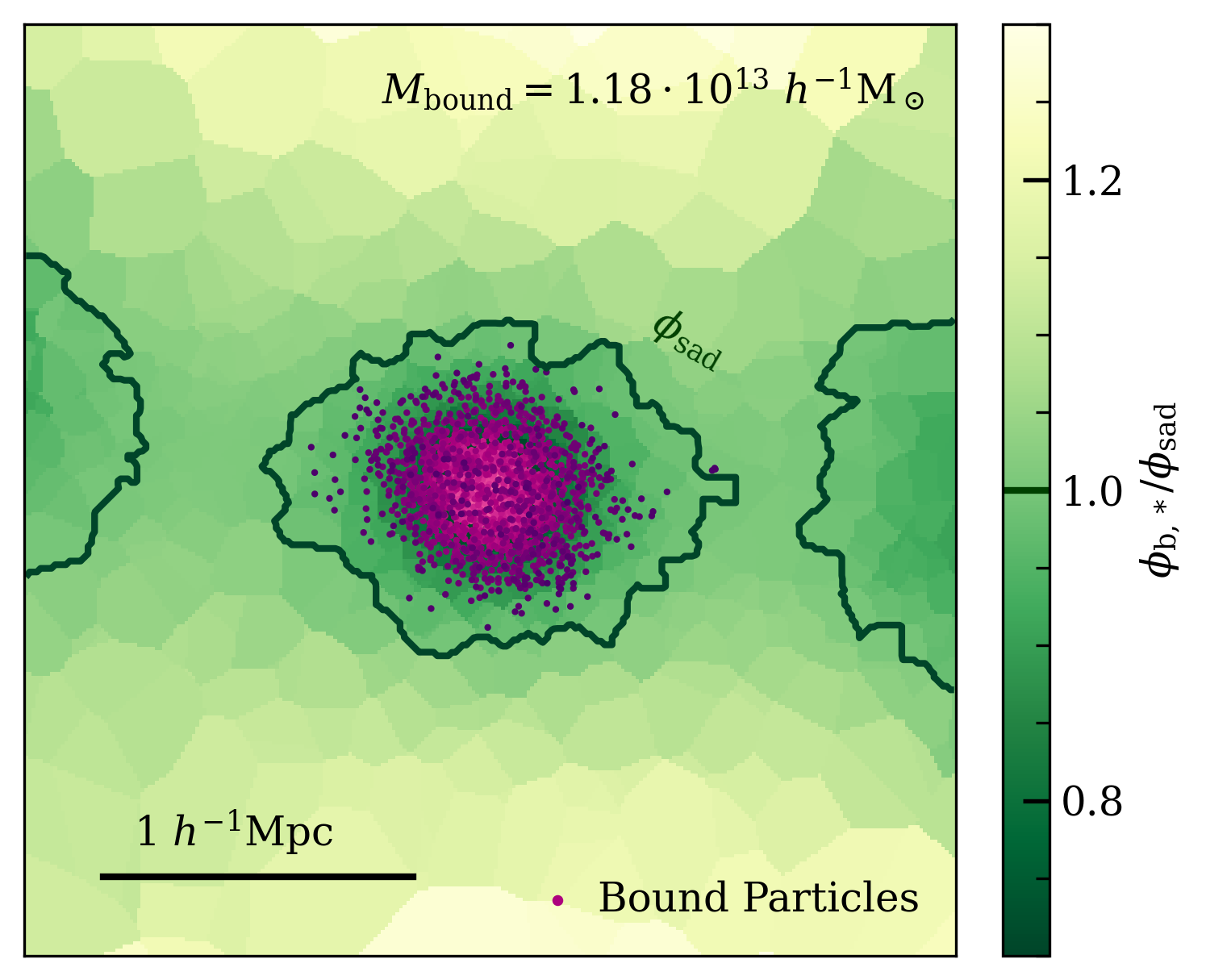}
    \caption{Slice of the boosted gravitational potential field centred on a halo, the dark green contours mark the saddle point energy, $\phi_{\rm sad}$, the bound population of particles is shown as magenta points.}
    \label{fig:single_halo_strip}
\end{figure}

 In $N$-body simulations, another popular choice of definition is the mass of Friends-of-Friends (FoF) groups, \citep[e.g.][]{Press1982,Subfind2001,Roy2014}, which depend on a linking length between individual particles that serves as a proxy for the local matter density. This has the advantage of removing the assumption of spherical symmetry, and being numerically efficient to implement. Moreover, this definition provides valuable insight as to the clumpy nature of haloes but comes at the expense of being more difficult to link to theory and not being directly applicable to observations. In addition, FoF has the tendency to detect chance alignment of particles, marking the presence of a halo where there is none, for example in \citet{Angulo2013} and \citet{Stuecker2020,Stucker2021} caustics and numerical fragments are detected by FoF as haloes in warm dark matter simulations. Furthermore, because FoF relies on the local density of particles, it may create unphysically large structures by bridging between haloes \citep[see e.g.][]{Knebe2011,Leroy2021}. This particular issue stems from the fact that haloes lack a sharp boundary
in terms of density. 

Nevertheless, in simulations the physical properties of dark matter, such as the density, velocity, and acceleration fields, are known. These allow to paint a more in depth picture of haloes by using more advanced definitions. While the edge of haloes is difficult to define from the density field alone, \citep[see e.g.][]{HOP1998,Subfind2001, Aubert2004,Codis2015,Codis2018}, it is possible to do so in terms of dynamical properties. For instance, one can select particles based on their orbital status, considering only particles that have undergone their first apocentric passage \citep{Adhikari2014,Sparta2017}, or which have undergone shell crossing in three dimensions \citep{Flack2012,Stuecker2020} as part of the halo. 
Alternatively, one can define the boundary in energy space by introducing cuts in terms of kinetic energy \citep{Garcia2023,Salazar2024}. These approaches are, however, much more computationally expensive, with the first methods requiring that one tracks the motion of every particle and the second requiring an internal fitting procedure to optimise the kinetic energy cut.

A quantity that is often overlooked when studying haloes is the gravitational potential, $\phi$.
As the scalar quantity sourcing the motion, the gravitational potential is the natural quantity defining the area of influence of a given halo. 
Until recently, using the gravitational potential has been impossible in practice since it is dominated by large-scale gradients which mask the potential wells of smaller structures. However, \cite{Stucker2021} showed that it is possible to use the gravitational potential to define structures by adding a `boost', 
\begin{equation}
    \phi_{\rm b}(\vec{x}) = \phi(\vec{x}) - (\vec{x} - \vec{x}_0) \cdot \left.\nabla_{\vec{x}} \phi(\vec{x})\right|_{\vec{x} = \vec{x}_0} = \phi(\vec{x}) + (\vec{x} - \vec{x}_0) \cdot \vec{a}_0,
    \label{eq:bopo}
\end{equation}
that removes the locally-averaged large-scale gradient, $\left.\nabla_{\vec{x}} \phi(\vec{x})\right|_{\vec{x} = \vec{x}_0}$, an operation equivalent to studying a halo in a reference frame free-falling with an acceleration $\vec{a}_0$. With the equivalence principle implying that this type of transformation does not impact the internal dynamics of the system. In this context, haloes can be defined as host potential wells, for which the boundary is set by a saddle point in the field which leads to a deeper potential well.
Particles within the well that have energies below the saddle-point energy are bound to the structure, whereas particles with higher energies could escape. This definition of boundness can be seen as a generalization of commonly used self-binding checks, \citep[e.g.][]{Subfind2001,Behroozi2013} since it naturally includes the gravitational effect of surrounding structures. This is illustrated in Fig.~\ref{fig:single_halo_strip}, where we show a slice of the boosted gravitational potential field centred on a $10^{13}\hMsun$ halo. Here we outline the saddle point energy, $\phi_{\rm sad}$, marking the dynamical edge of the halo, with a thick green line and show the particles which are bound to this potential well in magenta.

In this work, we propose to expand the study of the properties of haloes as defined using the boosted potential in \citet{Stucker2021}. As such, we develop a particle assignment algorithm, known as {\sc strawberry}\footnote{\textsc{Str}ucture \textsc{a}ssignment \textsc{w}ithin \textsc{b}oost\textsc{e}d \textsc{r}eference f\textsc{r}ames in c\textsc{y}thon}
, to efficiently find the saddle point energy and check whether particles are bound or not to a given potential well. 
This work is structured as follows. After discussing the theoretical framework in Sect.~\ref{sec:theory}, we will present our particle assignment algorithm in Sect.~\ref{sec:algorithm} for which several convergence tests can be found in App.~\ref{app:scaling_convergence}. In Sect.~\ref{sec:halo_props}, we study the properties of particle distributions assigned to haloes, from how particle energies evolve to how they are then distributed in the halo, the virialisation state, and compare bound masses of haloes to the more commonly used $M_{\rm 200b}$. Finally, in Sect.~\ref{sec:discussion}, we summarise this work and present our conclusions and future prospects.

\section{Theory} \label{sec:theory}
The goal of the {\sc strawberry} algorithm is to find bound structures in the gravitational potential. However, notions of boundedness and the potential energy, depend on the choice of a coordinate frame. Here, we will discuss our notion of boundedness, how it operates in the idealized case of a static potential landscape, how the expansion of the universe can be accounted for and how to choose an optimal frame of reference.

\subsection{Notion of boundedness} \label{sec:boundedness}
It is difficult to give a clear definition of what being `bound' means in the case of cosmological simulations. For isolated systems, binding notions tend to reference the gravitational energy normalized to $\phi(\rightarrow \infty) = 0$ and check whether a given particle's energy is sufficient to escape to infinity. However, this assumption shows immediate weakness when the additional potential of other structures is considered. For example, consider the gravitational potential of the Earth. In principle, it would be possible to bind particles at arbitrary large distances if they have low enough relative velocities. However, due to the gravitational field of the Sun, any particles at a distance greater than the Lagrangian point L1 will orbit around the Sun, rather than the Earth. While the self-potential binding check is a powerful heuristic in many situations, we probably do not want to use it as a baseline of what `being bound' means.

Intuitively, what we mean by particles being `bound' to a structure is that they are restricted to orbiting around it. Approximately, we may say that a particle $i$ is bound to a given set of particles at a time $t$, if for all future times it may not leave a certain area of influence. This notion of boundedness is however not very precise, as it has several ambiguities: 
(1) The coordinate frame that we use to define distances needs to be specified. For example, in cosmology, under this definition a structure may be bound in comoving space, but unbound in physical space.
(2) It is not clear which reference sets of particles and which area of influence should be considered. 
(3) It is also a somewhat impractical notion, as it requires predicting the full future of the particle.
For instance, if a particle orbits in the vicinity of a structure for a long time, but then gets expelled, e.g. through some event that strongly changes its energy -- it is not obvious whether or not it should be called `bound' before the ejection event.

Despite these ambiguities, we argue that the goal of any physical definition of boundness is to approximate statements of this form. That is, heuristics based on energy arguments which may be used to make approximate predictions about the future of particles. For a given heuristic, a particle then becomes bound when it becomes clear that it is restricted in this sense for all future times. Our goal is to define a heuristic based on the `full' gravitational potential that allows to detect bound structures in this sense.

\begin{figure}
    \centering
    \includegraphics[width=\linewidth]{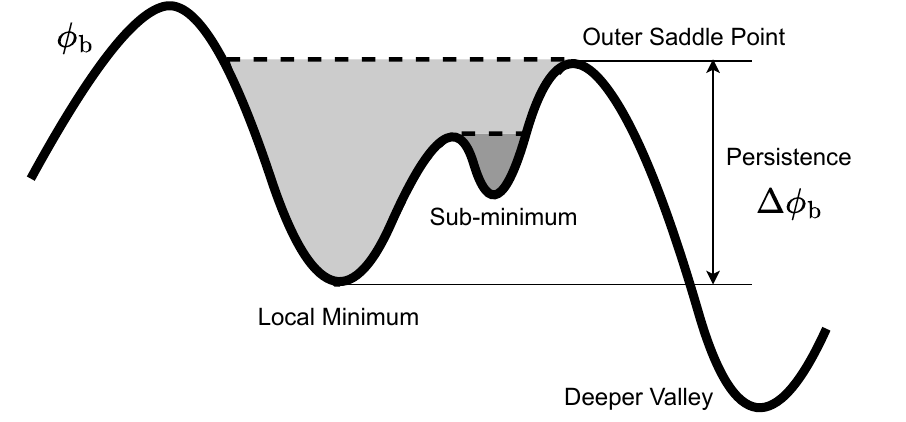}
    \caption{Illustration of our binding notion for a fixed potential landscape. The region where  particles may be bound to the minimum extends up to the first saddle-point that connects to a deeper minimum.}
    \label{fig:well_diagram}
\end{figure}

Before we discuss the full cosmological scenario, it is insightful to consider a simplified thought experiment. Let us assume we have a Hamiltonian of the form
\begin{align}
    H = \phi(\vec{x}) + \frac{1}{2}\vec{v}^2,
\end{align}
with a complicated (non-monotonic), but static potential $\phi$. As energy is conserved in this scenario, it is easy to find bound regions: If a particle has an energy level $E=H$, then it may only orbit in the space restricted by $\phi(\vec{x}) < E$. Further, it is restricted to the connected space (that contains its starting position) that fulfils this criterion. We may associate a connected region $\phi(\vec{x}) < E$ with the deepest minimum $\phi_{\mathrm{min}}$ inside of it. The maximal extent of space that may be associated with a given minimum is then limited by the first saddle-point $\phi_{\mathrm{sad}}$ that connects to a deeper minimum. The difference in energy $\Delta \phi = \phi_{\mathrm{sad}} - \phi_{\mathrm{min}}$ is called the `persistence'. Therefore, in the described scenario, we may say that a particle is bound to the deepest minimum in the connected space $\phi(\vec{x}) \leq E$. 
This is illustrated in Figure \ref{fig:well_diagram}.

While in the previous scenario we can therefore provide a very clear notion of boundedness, the cosmological scenario is considerably more complicated. This is because the potential notion depends on the adopted reference frame and is generally highly time-dependent. However, it is possible to recover a scenario that is close to the one described above, by transforming into an appropriate reference frame accounting for the expansion of the universe and for large scale accelerations. 

\subsection{Potential Energy}
In physical space, it is common to define the potential through the Poisson equation
\begin{align}
    \nabla_r^2 \phi_r &= 4 \pi G \rho(\vec{r}),
\end{align}
where $\nabla_r$ denotes a derivative with respect to physical coordinates $\vec{r}$ and $\rho(\vec{r})$ is the physical density. Physical energy is given by
\begin{align}
    E &= \phi_r(\vec{r}) + \frac{1}{2} v^2 - \frac{1}{6} \Lambda c^2 r^2, \\
      &= \phi_r(\vec{r}) + \frac{1}{2} v^2 - \frac{1}{2} H^2 \Omega_{\Lambda}(a) r^2,
\end{align}
where $\Lambda$ is the cosmological constant. The dark energy term is often omitted, as it is usually sufficiently small at the small distances that are of interest in physical space. However, since it has an effect on larger scales, it is included for consistency.

In the comoving frame, we may write
\begin{align}
    \vec{x} &= \frac{\vec{r} }{a}, \\
    \vec{v} &= \vec{\dot{x}} a^2 = a ( \vec{\dot{r}} - \vec{r} H), \\
    \nabla_x^2 \phi &= 4 \pi G a^2 [\rho(\vec{r}) - \rho_{\rm m}] = \frac{3}{2} a^{-1} \Omega_{\rm m,0} H_0^2 \delta, \\
    \vec{\dot{v}} &= - \nabla_x \phi,
\end{align}
where $x$ is the comoving coordinate, $a$ is the scale factor, $\vec{v}$ is the canonical momentum which relates to peculiar velocities as $\vec{v}_{\mathrm{pec}} = \vec{v} / a$, $\rho_{\rm m}$ is the mean matter density of the universe, $H_0$ the Hubble parameter at $a=1$, $\delta$ the relative matter overdensity, $\phi$ is the peculiar potential, and the dot marks derivatives with respect to the proper time for a comoving observer. Note that conventions on the definition of the peculiar potential may differ by a factor $a$, for example the cosmological simulation code Gadget \citep{springel_2021} outputs  $\phi_{\rm gadget} = a \cdot \phi$ in snapshots. Importantly, the physical potential and the peculiar potential differ by a quadratic term
\begin{align}
    \phi &= \phi_r - \frac{1}{4} H^2 r^2 \Omega_{\rm m},
    \label{eq:pot_transform}
\end{align}
which accounts for the effect of the mean density. We can express the physical energy from the comoving frame:
\begin{align}
    E &= \frac{1}{4} H^2 r^2 [\Omega_{\rm m}(a) - 2 \Omega_{\Lambda}(a)] + \phi + \frac{\dot{r}^2}{2}, \nonumber \\
        &=\frac{1}{4} H^2 a^2 x^2 \left[2 + \Omega_{\rm m}(a) - 2 \Omega_{\Lambda}(a)  \right] + H \vec{v}\cdot \vec{x} + \phi + \frac{v^2}{2a^2}, \nonumber \\
        &= \frac{3}{4} H^2 a^2 x^2 \Omega_{\rm m}(a) + H \vec{v}\cdot \vec{x} + \phi + \frac{v^2}{2 a^2},
\end{align}
where the third line is assuming a flat universe without radiation, so $\Omega_{\Lambda} = 1 - \Omega_{\rm m}$. 
We can calculate $E$ from the two different frames and get consistent results for the same particle. However, the split between which part of the energy is considered as potential energy and kinetic energy is frame-dependent. For a binding check, this implies that we may define a procedure which is consistent between resting and expanding reference frames by requiring $E < E_\infty$  where $E_\infty$ is some reference energy above which a particle may escape the potential well. The motivation of using the physical energy $E$ is that it is closest to a conserved quantity for static physical systems that have decoupled from the expansion of the Universe. 

The remaining challenge is to define $E_{\infty}$. In the scenario considered in Section \ref{sec:boundedness}, it appeared natural to consider saddle-points of the potential as the largest energies and furthest points where particles may be bound. However, due to the quadratic term in Eq.~(\ref{eq:pot_transform}) we have different saddle-points in the comoving and the physical frame. In absence of dark energy, saddle points of the physical potential signify points that would stay at a fixed physical distance if $\dot{r} = 0$ whereas saddle-points of the peculiar potential signify points that stay at a fixed comoving distance if $v = 0$. This ambiguity is similar to the ambiguity in the notion of boundedness that we have pointed out earlier.

A further necessary consideration is that it is easy to find examples where, for both the comoving and physical potential,
there is no saddle point.
For instance, let us take an overdensity $\delta(\vec{r})$ which, when spherically averaged within a radius $r$, is monotonically decreasing and positive, $\overline{\delta}(r) \geq 0$, for all $r$, and $\overline{\delta}(r \rightarrow \infty ) \rightarrow 0$. If we compute the physical potential,
\begin{align}
    \partial_r \phi_r &= \frac{4 \pi G}{3} r \rho_{\rm m} [1 + \overline{\delta}(r)],
\end{align}
we find that it is monotonically increasing ($\partial_r \phi_r$ > 0), and, as such, has no saddle point. In contrast, for the peculiar potential, it is
\begin{align}
    \partial_r \phi &= \frac{4 \pi G}{3} r \rho_{\rm m} \overline{\delta}(r), 
\end{align}
which reaches $\partial_r \phi \rightarrow 0$ only at infinity. In practice, neither case is convenient.

Pragmatically, we must therefore refine our definition and ask what is the furthest point that we may want to consider in a binding check? To answer this question, we can call upon the spherical collapse formalism. In this context, we may propose that we do not want to consider in our binding check any points beyond the turn-around radius, $r_{\rm ta}$, where spherically collapsing shells have $\dot{r} = 0$. Within this formalism, this radius corresponds to an average enclosed density $\tilde\delta$, with $\tilde{\delta} \simeq 4.55$ for an Einstein-de-Sitter (EdS) universe \citep[see e.g.][]{MBW2010}. From these ingredients, we construct a potential
\begin{align}
    \partial_r \phi_* =  \frac{4 \pi G \rho_{\rm m}}{3} r [\overline{\delta}(r) - \tilde{\delta}],
\end{align}
explicitly forcing the existence of a saddle point at  $\overline{\delta}(r) = \tilde{\delta}$. We refer to this potential as the turn-around potential, which can further be expressed as 
\begin{align}
    \phi_* &= \phi - \frac{1}{4} H^2 \Omega_{\rm        m} r^2 \tilde{\delta}, \\
           &= \phi - \frac{4 \pi G}{6} \rho_{\rm m} r^2 \tilde{\delta}.
\end{align}
While this potential only guarantees for a spherically symmetric system that the saddle will be at the turn around radius, it does, however, guarantee that at sufficiently large radii a saddle-point and a deeper minimum exist even for rare pathological cases. 

In terms of this potential, the physical energy is given by
\begin{align}
    E = \frac{1}{4} H^2 r^2 [\Omega_{\rm m}(a) (1 + \tilde{\delta}) - 2 \Omega_{\Lambda}(a)] + \phi_* + \frac{\dot{r}^2}{2},\label{eqn:Ebstar}
\end{align}
and we define the reference energy, $E_{\infty}$, as the energy level of a particle that is physically at rest $\dot{r} = 0$ (therefore, $\vec{v} = - a^2 H \vec{x}$) at the saddle-point, $\vec{x}_{\rm sad,*}$,
\begin{equation}
    E_{\infty} = \frac{1}{4} H^2 a^2 x_{\rm sad*}^2 [\Omega_{\rm m}(a) (1 + \tilde{\delta})  - 2 \Omega_{\Lambda}(a)] + \phi_*(\vec{x}_{\rm sad*}).
    \label{eqn:E_sad}
\end{equation}

We note that the choice of the potential that is used to define the saddle-point energy level leaves a degree of freedom that may seem somewhat arbitrary. However, in practice the difference between the different potential notions is almost always negligible, since they only tend to differ at much larger radii than the location of the saddle-point. We will show this in more detail in App.~\ref{app:dep_delta}.

\subsection{The boosted potential}
Definitions of energy depend on the adopted reference frame. For example, a Galilean transformation 
changes the kinetic energy term and it modifies the time-dependence of the potential. Furthermore, a transformation into an accelerated reference system with an arbitrary time-dependent offset $\vec{x}_0(t)$,
\begin{align}
    \vec{x}' &= \vec{x} - \vec{x}_0(t),\\
    \vec{v}' &= \vec{v} - \vec{v}_0(t),
\end{align}
with $\vec{v}_0 = a^{2} \dot{\vec{x}}_0$,
introduces an additional fictitious force in the transformed coordinates:
\begin{align}
    \dot{\vec{v}}' &= \dot{\vec{v}} - \vec{a}_0(t),
\end{align}
with $\vec{a}_0(t) = \dot{\vec{v}}_0(t)$. With the equation of motion $\dot{\vec{v}} = -\nabla_x \phi$, we may interpret this as a modification of the potential,
\begin{align}
    \phi_{\mathrm{b}}(\vec{x}', t) &= \phi(\vec{x}(\vec{x}', t), t) + \vec{x}' \cdot\vec{a}_0(t),
\end{align}
which we call a `boosted potential' \citep{Stucker2021}. According to the equivalence principle, transformations of this type do not affect the internal dynamics of systems, but they do change the energies that we would define for particles. Therefore, we need to choose a unique reference frame to define energies. Here, we argue that the preferred reference frame is the one that minimizes the explicit time-dependence of the potential in a region of interest. In such a frame, the conservation of energies is the least violated and a binding check the most meaningful.

Since the potential arises from the mass distribution, the explicit time-dependence of the mass distribution, $\rho(\vec{x}', t)$, should be small in an optimal reference frame.
As such, given a set of particles $\{ \vec{x}_i, \vec{v}_i\}$ -- e.g. all particles in a region of interest -- we choose our reference frame so that the centre of mass, the centre of mass velocity, and the centre of mass acceleration are fixed at 0: 
\begin{align}
    \langle \vec{x}' \rangle (t) &= 0;\\
    \langle \vec{v}' \rangle (t) &= 0;\\
    \langle \vec{a}' \rangle (t) &= 0,
\end{align}
where the expectation values go over all considered particles. We can then evaluate the transformations in terms of our original coordinates:
\begin{align}
    \vec{x}' &= \vec{x} -\vec{x}_0(t) = \vec{x}  - \langle \vec{x} \rangle(t), \\
    \vec{v}' &= \vec{v} - \vec{v}_0(t) = \vec{v} - \langle \vec{v} \rangle(t), \\
    \vec{a}' &= \vec{a} - \vec{a}_0(t) = \vec{a}  + \langle \vec{\nabla} \phi(\vec{x}) \rangle(t).
\end{align}

Combining the definition of the boosted potential with that of the turn around potential, we define, 
\begin{equation}
    \phi_{\rm b,*} = \phi - \langle \nabla \phi \rangle\cdot \vec{x} - \frac{1}{4} H^2 \Omega_{\rm m} r^2 \tilde{\delta}.
    \label{eq:phi_b_star}
\end{equation}
in the accelerated reference frame. This potential is the field that will be used in the following section to define the dynamical boundary of haloes, and energies of particles by replacing $\phi_*$ with $\phi_{\rm b,*}$ in Eqs.~(\ref{eqn:Ebstar}) and (\ref{eqn:E_sad}).
Indeed, this definition grants us all of the benefits discussed above by both minimising the explicit time dependence of the potential while also guaranteeing that there will always exist a saddle point to which we can define the boundary of the potential well.

\section{Algorithm}
\label{sec:algorithm}

\begin{figure}
    \centering
    \includegraphics[width=\linewidth]{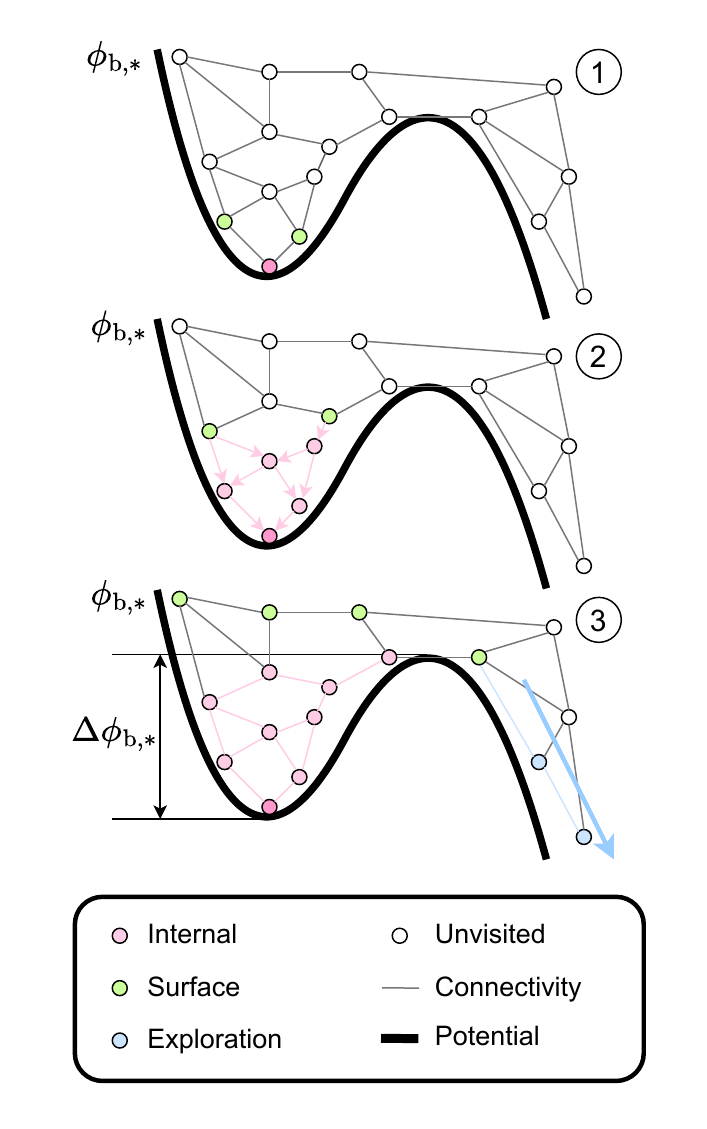}
    \caption{Illustration of main particle assignment scheme implemented in {\sc strawberry}. We start by defining the turn around boosted potential, $\phi_{\rm b,*}$, shown as a thick black line, and assigning the connectivity, thin grey lines, between the particles, drawn as circles. We then place ourselves at the bottom of the potential well and defining the group within the well, pink particle, and tracking particles adjacent to this group, in green, also known as surface particles. In the second panel, we grow the pink group by including surface particles if all of their neighbours with lower potentials are also part of the group. In the third panel we stop growing the group as we have found a surface particle, to the right, which is connected to particles, in blue, which lead to a deeper potential valley. This final particle sets the persistence of the group, $\Delta\phi_{\rm b, *}$. }
    \label{fig:assignement}
\end{figure}

In the previous section, we have laid down the theoretical framework upon which we construct our particle assignment scheme. In this section we go over more specifically how this framework is adapted to sets of particles and how it is implemented to produce groups of particles bound to potential wells. To summarise this section, the {\sc strawberry} halo assignment algorithm, hereafter simply referred to as the algorithm or {\sc strawberry}, proceeds in four main steps.
\begin{enumerate}
    \item Propose a region of interest
    \item Apply the transformation to an accelerated frame.
    \item Find the saddle-point contour of the boosted turn-around potential, $\phi_{\rm b,*}$
    \item Perform the binding check by evaluating whether $E < E_{\infty}$ for all particles within the saddle-point contour. 
\end{enumerate}

\subsection{Moving about the simulation}

Before getting into the core of the algorithm, one crucial step is however required. That is, to translate the theoretical framework of Sect.~\ref{sec:theory} which is expressed in terms of continuous fields, into a form applicable to a discrete set of particles. In practice, this step is important because we rely on features of these fields to define structures, namely minima, gradients, and saddle points, making them an integral part of the numerical scheme. As will be discussed further in Sect.~\ref{sec:fill_well}, we construct structures by iteratively adding neighbouring particles until we fill the potential well. This requires two elements, first a means of computing the boosted potential at the location of each visited particle, and second a definition of neighbouring particles.

As such, {\sc strawberry} begins by establishing a connectivity between particles. In this context, connectivity can be seen as a set of edges which connect particles to each other, assigning to each one a set of neighbours. There are many ways that one could define this connectivity. For instance, a common choice is to use a Delaunay tessellation \citep{forman_1998, forman_2002, sousbie_2011, tierny2018,Stucker2021}, which has many desirable properties, such as guaranteeing that any particle is accessible from any other particle in the simulation while using a minimal number of edges. This kind of connectivity is however computationally expensive, making cheaper alternative more desirable given the large number of particles used in modern cosmological simulations.

Here, we instead opt for a simpler connectivity definition. For each particle, the algorithm saves a list of the $N_{\rm ngb}$ nearest neighbours, with {\sc strawberry} tracking by default the ten nearest neighbours of each particle. This choice is both simple to implement, due to the fixed size of the list of neighbours, and is efficient to compute, through the use of the $k$-d tree algorithm \citep{Bentley1975} for example. Because this particular connectivity does not have the guarantees of more advanced choices, it can be prone to forming clumps and islands which prohibit the exploration of certain parts of the simulation. However this can be easily avoided if the number of saved neighbours is sufficiently high, see App.~\ref{app:neighbours} where we show that for the number of neighbours used by default, the recovered physical properties of haloes are converged.
Alternatively, one may also provide information about the field that is being explored, for example {\sc subfind} \citep{Subfind2001} establishes an effective connectivity based on the density field by connecting each particle to its two neighbours with highest density within a preselected list of nearest neighbours. While this may seem more efficient, in the case at hand this would mean recomputing the connectivity each time we switch reference frame, resulting in a significantly increased computational cost.

\subsection{Initialisation and reference frame}
\label{sec:init_guess}
As stated previously, the definition of the boosted potential relies on local averages in order to pass into the accelerated reference frame of the structure. Without an initial estimate of the positions and sizes of structures it is difficult to define these quantities \citep[see e.g.][]{Stucker2021}. 
As a first guess, our algorithm uses an initial catalogue of particles split into FoF groups. It is from these groups that we compute mean acceleration, $\langle \vec{a}\rangle$, along with the mean position of particles, $\langle \vec{x}\rangle$, which provides an initial guess as to the location of the minimum of the potential well.  We note that these FoF groups only serve as a selection of the zones of interest that the {\sc strawberry} algorithm will visit, and are not used during the assignment and binding procedures described in Sect.~\ref{sec:fill_well}.

Using this initial guess we compute the boosted potential for all particles within the FoF group and select the particle which has the lowest potential value. In the majority of cases, over 99 percent, the resulting particle has no neighbours with lower potential values, meaning that it resides at the bottom of a potential well.
In the remaining cases, which are primarily highly perturbed haloes or numerical artefacts, the selected particle does not fulfil this condition, typically residing on a slope in the potential at the boundary of the FoF group. 
In these cases we perform an approximate gradient decent starting from the mean position of particles in the FoF group, which finds the minimum of the well. In the rare cases where this second procedure does not find a minimum and instead leads to particles that are not within the FoF group, fewer than 0.01 percent of haloes, then the FoF halo is discarded. 

\subsection{Filling the potential well}
\label{sec:fill_well}

With all of this in hand, we can now fill the boosted potential well and locate the saddle point. We achieve this by defining a connected group of `internal particles', and continuously  keeping track of a set of `surface particles' which are all the neighbours of the internal group that are not marked as internal particles. We respectively initialise these two groups as containing only the potential minimum and all of its neighbours.
To grow the group, and fill the potential well, the assignment procedure transfers surface particles to the internal group until the saddle point is found, after which the binding procedure is performed on the internal particles.
We have illustrated this process in Fig.~\ref{fig:assignement}, where we take a simple one dimensional example for ease of comprehension. 

In the second phase, panel 2, we grow the internal group by iteratively including surface particles. To do so, we select the surface particle which has the lowest potential value, and include it in the group if all its neighbours with lower potential values are also part of the internal group. If the particle is added to the growing structure, all of its unassigned neighbours are now considered surface particles. 

We repeat this step until we find a surface particle which does not satisfy this condition, i.e. which has neighbours with lower potentials which are not part of the internal group.
This configuration can occur in two cases, either we have found the saddle point which leads to a deeper potential valley, 
or we have found a substructure, as is illustrated in Fig.~\ref{fig:well_diagram}. To check if we have found the saddle point which leads to a deeper potential valley, we create a new exploratory subgroup, blue particles in Fig.~\ref{fig:assignement}. We explore this region by repeatedly connecting the lowest potential exploration-group-adjacent unvisited particle to this subgroup.
This procedure continues until we either reach a particle which has a boosted potential value that is lower than the minimum, or we run out of particles that have potential values lower than the candidate saddle point.
If we encounter a particle that has a lower potential than the minimum of our well, then we consider that we have found the desired saddle point and have thus filled the well. In this case we then define the persistence, $\Delta\phi_{\rm b,*}$, as the difference in the potential between this saddle point and the lowest point in the well. If not, we are in the presence of a sub-minimum. All the selected particles are then included into the internal group and all of their unassigned neighbours tracked as surface particles. Note that we keep track of these subgroups of particles as they require special attention later during the binding check. 

\subsection{Binding check}

The final step of this assignment pipeline, is the binding check. The energy of every selected particle is computed using Eq.~(\ref{eqn:Ebstar}) and is then compared to the energy at the saddle point given by Eq.~(\ref{eqn:E_sad}), after recentring the reference frame around the minimum of the potential well. Particles which have an energy lower than the saddle point energy are considered bound, and those above are consider unbound, thus creating two populations of particles, bound and unbound, inhabiting the potential well. From this selection, we define the bound mass, $M_{\rm bound}$, of the halo as the total mass of particles assigned to the bound group.

In most dark matter models, haloes present a certain amount of substructure in the form of subhaloes. These objects, just like their hosts, are gravitationally bound, and influence the potential landscape by creating local minima, which are in motion within the potential of the host halo. When it comes to the binding procedure, these local minima are problematic because, if we do not account for their motion, they allow certain particles to pass the binding check which would have otherwise failed. An intuitive example of this is that of a substructure that is being accreted onto the central halo. Let us consider a particle within the substructure which is, by chance, moving in exactly the opposite direction to the bulk motion of the substructure. In the reference frame of the central halo, its peculiar velocity is then ${\bf v} = {\bf 0}$. In this case, the energy of the particle is simply given by the sum of the potential of the central halo and infalling halo, along with the Hubble term. Given this configuration and provided that the infalling structure is sufficiently close to the potential minimum, the lack of apparent motion will allow the particle to easily pass the binding check. To avoid this issue, we introduce a secondary bulk binding check which is applied to substructures.

Before doing so we must first decide which particles should be considered part of a subgroup. In practice, this equates to simply analysing the local potential minimum in the same manner as the main group. We recall that, during the group assignment procedure, we assigned certain particles to candidate subgroups for being in the vicinity of a local dip in the boosted potential. In the case where the subgroup is sufficiently large\footnote{By default 20 particles}, it is likely that the local minimum in the potential is generated by the presence of a substructure. 

After the central halo's binding energy is found, each subgroup is re-analysed, doing so in the same way we analysed the main group, using each candidate subgroup as a seed for a substructure in the same way we used a FoF group for the main halo. Using these subgroup particles, regardless of whether they are bound to the main structure or not, we measure the local average acceleration, find the particle that has the lowest potential in this sub-well, and finally repeat the assignment and binding procedures. From this we obtain a new subgroup of particles that are gravitationally bound to the local minimum. 

We treat the binding of the whole subgroup to the host halo as a single object, with its kinetic energy being defined by the bulk velocity of the structure and using its maximum potential, in the reference frame of the main structure, to estimate its potential energy. Given these quantities, we either bind or discard the entire structure, in the same fashion as individual particles in the main group.

\section{The bound population of haloes}
\label{sec:halo_props}

The \textsc{strawberry} algorithm presented in the previous sections allows us to study large quantities of haloes through the rapid analysis of simulation snapshots, and producing catalogues of halo properties. In this section, we analyse the resulting particle distributions to better understand the bound population. Specifically, we first focus on where and when particles become bound to their host, how these particles are then distributed within the halo in terms of energies, and how this selection translates to positions and velocities. Here, we focus on halo dynamics in a single $\Lambda$CDM cosmology and leave a larger analysis of any possible cosmological dependence for future work. 

To perform our study, we use a gravity-only N-body $\Lambda$CDM simulation, for which the cosmological parameters are set to the values of the Illustris TNG300 simulation \citep{TNG2018}, namely, $\Omega_{\rm m,0} = 0.3089$, $\Omega_{\Lambda,0} = 0.6911$, $\Omega_{\rm b,0} = 0.045$, $n_{\rm s} = 0.9667$, and $h = 0.6774$. The simulation is run using the \textsc{gadget 4} \citep{Gadget4_2021} code and represents a comoving periodic box of side length $L_{\rm box} = 205 h^{-1}{\rm Mpc}$ sampled by $N=625^3$ particles, leading to a particle mass of $3\cdot10^9\hMsun$. The initial conditions are generated at $z_{\rm init} = 99$ using second-order Lagrangian perturbation theory as implemented in the built-in initial condition generator, \textsc{ngenic} \citep{NGENIC}, using the \citet{Eisenstein1998} approximation for the linear power spectrum.

\subsection{When and where do particles become bound to a halo?}

Asking ourselves the question, `when and where do particles become bound to haloes?' echoes dynamical halo boundary definitions, where the boundary of haloes is set not in terms of density but the dynamics of particles such as particle orbits or energies \citep{Sparta2017,Garcia2023,Salazar2024}. 
To answer this question, we select a sample of 100 haloes with bound masses between $5\cdot10^{13}\ \hMsun$ and $1.5\cdot10^{14}\ \hMsun$ for which we track the assembly history from $a=0.5$ to $a=1$, and simultaneously track all particles which enter within a spherical region of radius $R<1.5\sqrt{3}\simeq2.6\ h^{-1}{\rm Mpc}$ centred on the position of the main progenitors. During the tracking procedure we travel backwards in time defining the main progenitor group as the FoF halo within the tracked region which contains the most particles which will be bound to the main descendant halo in the following snapshot. Once the main progenitor is detected, we recentre the region on its location before proceeding with the particle assignment procedure described in Sect.~\ref{sec:algorithm}. Within each region we track the individual positions, velocities, boosted potential values, and energies of all particles.

\begin{figure}
    \centering
    \includegraphics[width=0.95\linewidth]{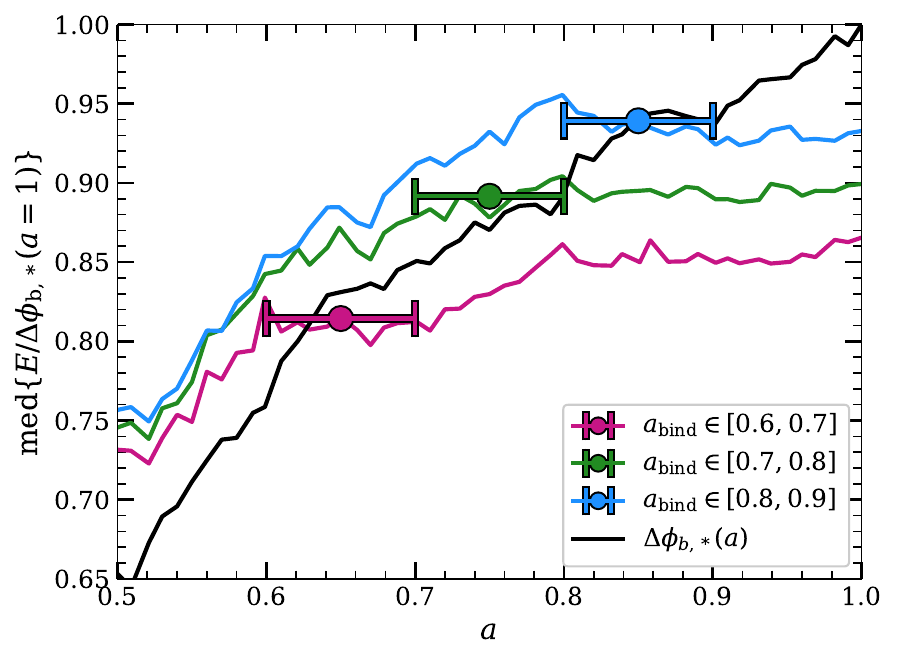}
    \caption{Median energy histories of three sets of particles from a sample of 100 haloes with masses $M_{\rm bound}\in[0.5\cdot10^{14}\hMsun,1.5\cdot 10^{14}\hMsun]$, selected according to their respective time of binding.
    The individual energy histories of particles from different haloes are normalised by the persistence $\Delta\phi_{\rm b}(a = 1)$ of the host halo before the median is computed. The median persistence history of the sample of haloes is shown in black.}
    \label{fig:energy_tracks}
\end{figure}

From the resulting dataset, we estimate the energy histories of all particles that enter theses 100 regions. Among these energy tracks, we select three subsamples of particles which become bound to their host haloes at scale factors $a \in [0.6;0.7]$, $a\in[0.7,0.8]$, and $a\in[0.8,0.9]$. In order to stack energy histories originating from different haloes, we normalise the individual histories by the persistence of their host halo at $a=1$.  
In Fig.~\ref{fig:energy_tracks} we show the evolution of the median energies of all three subsamples along with the evolution of the median persistence, normalised at $a=1$.
Over time, the energy of particles continuously grows slowly by about $20\%$ to $30\%$ between $a=0.5$ and $a=1.0$ up to the time when they are accreted, after which their energy stays constant. On the other hand, the persistence grows at a significantly faster rate, almost doubling since $a=0.5$. Since particles become bound when they cross the persistence threshold, the core reason that particles become bound here is that the persistence grows -- primarily by additional mass being accreted outside of the orbital radius of the particles of interest. 

Note that this is quite different than what would be observed if the self-potential was used to define energies. A notable difference is that the self-potential is normalized to  zero at infinity, whereas our potential notion is normalized to be zero at the minimum. In the self-potential, particles become bound because they decrease their self-energy and cross the zero-energy level. E.g. if material is accreted at far larger radii than the orbital radius of a particle of interest, then the self-energy decreases, whereas the energy defined here would be unaffected (while the persistence of the potential increases). This can be seen in Fig.~\ref{fig:energy_tracks} where the energy history of particles flattens out after they become bound. We argue that this is a desirable property, and suggest that normalizing energies to zero at the minimum may make comparisons of energy levels across time more meaningful. Indeed, we can see that the energies remain clearly separated and maintain their ordering after accretion, indicating that once the halo is virialised it maintains a memory of the accretion process. Note that such a perspective also simplifies understanding other processes, like tidal stripping \citep[e.g.][]{Darkos2020,Errani2022,Stuecker2023}.

\begin{figure}
    \centering
    \includegraphics[width = 0.95\linewidth]{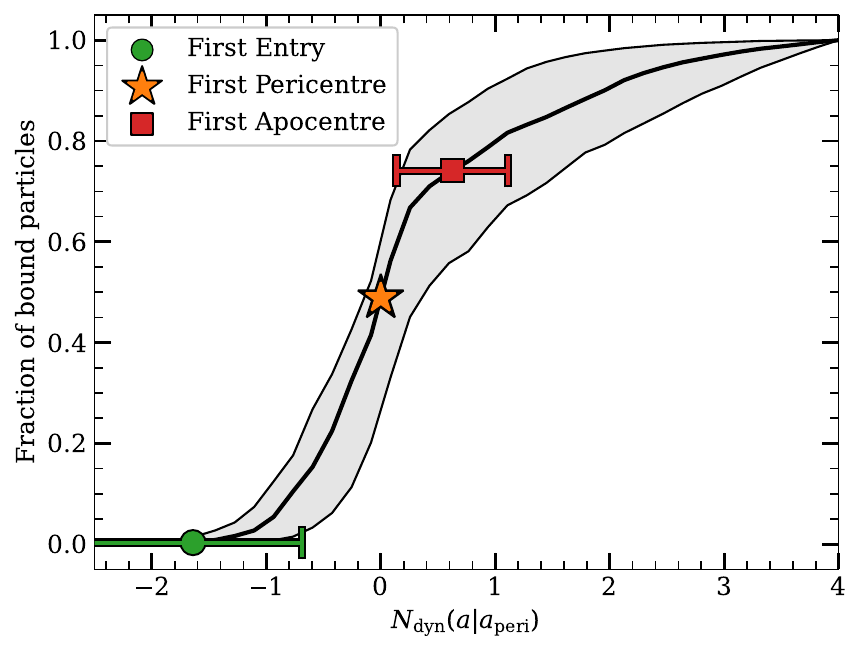}
    \caption{Fraction of bound particles as a function of the number of dynamical times since the first pericentric passage, shown as the median, (solid black line) and 1-$\sigma$ region (shaded area) as estimated over 100 haloes. The coloured points represent different events in a particle's history, green, the particle enters the potential well, orange, the particle passes the pericentre of its orbit for the first time, and red, the particle reaches its first apocentre.}
    \label{fig:binding_prob}
\end{figure}

To gauge at what point of their orbit particles become bound, in Fig.~\ref{fig:binding_prob} we compute the fraction of bound particles after a given number of dynamical times \citep{Jiang2016},
\begin{equation}
    N_{\rm dyn}(a|a_{\rm peri}) = \frac{\sqrt{2}}{\pi}\int_{a_{\rm peri}}^{a}\frac{1}{a}\sqrt{\frac{\bar{\rho}}{\rho_{\rm c}}}\dd a,
\end{equation}
since the first pericentric passage, with $\bar{\rho}$, the characteristic density of haloes which we choose to be 200 times the mean cosmological density, $\rho_{\rm c}$, the critical density, $a$, the scale factor, and, $a_{\rm peri}$, the scale factor at first pericentric passage. The pericentre is chosen as the first minimum radius with respect to the position of the potential minimum, and maximum velocity norm with respect to the average velocity of the bound population.
This specific choice of variable allows us to stack the binding histories of all particles, regardless of the time of their first pericentric passage. 

We find that the fraction of bound particles begins to increase as soon as particles enter the saddle point contour of the halo, 
we mark this moment with a green circle, with the error bars around this point marking the standard deviation of $N_{\rm dyn}$ between the particle entering the halo and its first pericentric passage.
Most particles that become bound at this stage are particles that are accreted with lower-than-average energies. 
As we approach the pericentre, the fraction of bound particles increases rapidly, with 50\% of particles being bound to the structure at the first pericentric passage, marked by an orange star\footnote{This point is by definition at $N_{\rm dyn} = 0$ and hence does not have any dispersion.}. At the first apocentre, marked by a red square, we find that 75\% of particles are considered bound to the structure. Finally, we observe that after two dynamical times, 90\% of the particles are considered bound. From these considerations, we confirm that gravitational binding occurs on a relatively short timescale, typically only a few dynamical times. 
These observations support the motivations behind particle selection criteria based on orbital status.  
For instance, one may select particles based on whether it has passed its first pericentre or apocentre \citep[e.g][]{Adhikari2014, Sparta2017}, meaning parallels can be drawn between these types of mass definition and the one presented in this work.

\begin{figure}
    \centering
    \includegraphics[width=0.95\linewidth]{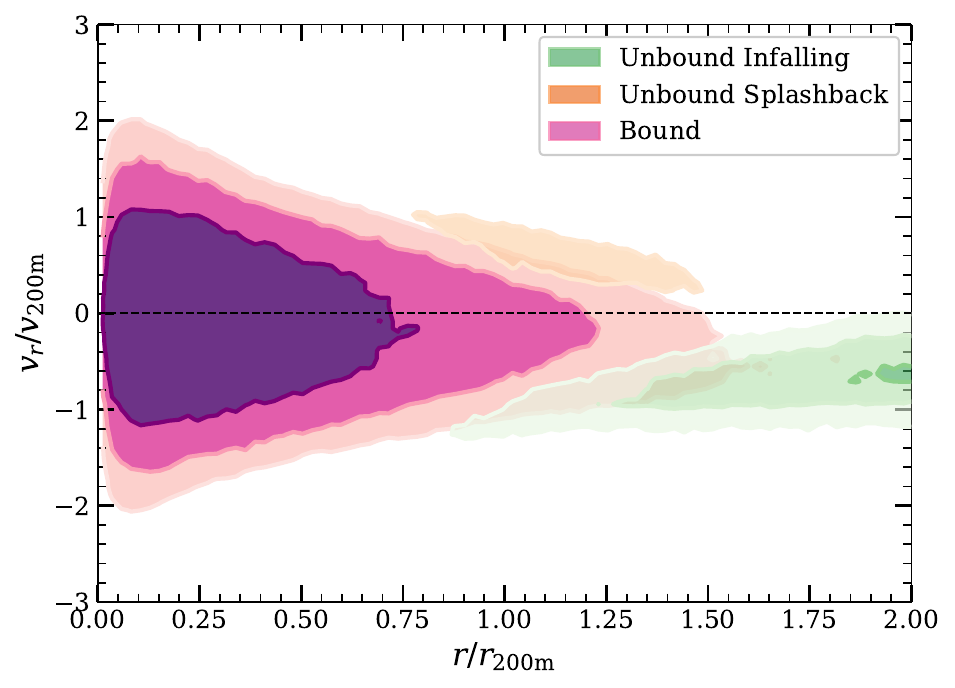}
    \caption{Stacked radial phase space density distribution of 100 massive haloes in the simulation. In magenta, we show the bound population and in green the infalling population. From darkest to brightest, both contours jointly encompass, 50\%, 84\%, and 95\% of all particles. We note that the splashback contour only displays one shade.}
    \label{fig:phasespace_100MMH}
\end{figure}

The binding of particles can further be explored in phase space. Indeed, it is generally accepted that in phase space, CDM haloes can be represented as ravelled three-dimensional manifolds \citep{Shandarin1989}. When projected in terms of radius, $r$, and radial velocity, $v_{\rm r}$, haloes appear as a spiral pattern, with the most recently accreted material being deposited on the outskirts of the halo and older material being concentrated in the centre. The halo is then connected to the background medium through a stream of infalling material \citep[see e.g.][]{Wang2011, Adhikari2014, Zavala2019}.

In Fig.~\ref{fig:phasespace_100MMH}, we show the stacked radial phase space density of the same 100 haloes, 
The phase space density is split into bound (magenta), unbound and infalling (green), and unbound and splashback (orange) populations. We select bound particles according to the boosted potential criterion presented above and separate the unbound sets using the orbits of particles. The brightness of each contour is set to enclose 50\%, 84\%, and 95\%, respectively from darkest to lightest, of the total population, including both bound and unbound population. 
Here, the bound population occupies a triangular region around the centre of the halo. This population is relatively symmetric in terms of radial velocity. As such it is reasonable to think that this population is virialised, we discuss this quantitatively in Sect.~\ref{sec:virial}. This region of phase space, is flanked at larger radii and radial velocities by material classified as unbound. We observe that the second component can be split into two parts, one with negative radial velocities, which we refer to as the first infall stream and a second with positive radial velocities, which we refer to as the first splashback stream. We note that the first splashback stream is less dense than the first infall stream. This is in line with our previous observation, where we find that by the first pericentric passage approximately half of particles are already bound to the halo, as such they will not be included here. This configuration of bound and unbound particles is consistent with the kinetic energy - radius cuts proposed by \citet{Salazar2024} to filter out recently accreted material. Indeed, when plotted in terms of kinetic energy and radius, the distinction between bound and unbound particles in Fig.~\ref{fig:phasespace_100MMH}, follow the valley features used in \citet{Salazar2024} to define the orbiting and infalling populations. 

\subsection{Matter density profiles}
\label{sec:halo_profile}

Given the previous observations, let us now investigate in more detail how the bound population is distributed by studying the matter density profile of the resulting structures.
Through simulations, it has been shown that haloes on exhibit the same density profile over at least 20 orders of magnitude in halo mass \citep{wang2020_prof}. In particular, it has been shown that the density profile is well described by a Navarro-Frenk-White (NFW) profile \citep{NFW1997},
which depends solely on a single scale radius and amplitude. This profile, however, has the significant shortcoming of not having a convergent mass, meaning that integrating the profile to infinitely large radii does not converge to a finite value.
Because of this ambiguity over where haloes `end', it has become common practice to define mass with respect to an average enclosed overdensity $\Delta$, which has the advantage of being independent of the specific profile of the halo, thus standardising the definition of `halo mass'.
In recent years, this standardisation has been put into question. Indeed, using dynamical information \citep{Adhikari2014}, it has been shown that haloes, contrary to the most popular profiles used to describe them, have a finite extent, as shown by e.g. \citet{Sparta2017,Diemer2022}, \citet{Garcia2023}, and   \citet{Salazar2024}. 

\begin{figure}
    \centering
    \includegraphics[width=0.95\linewidth]{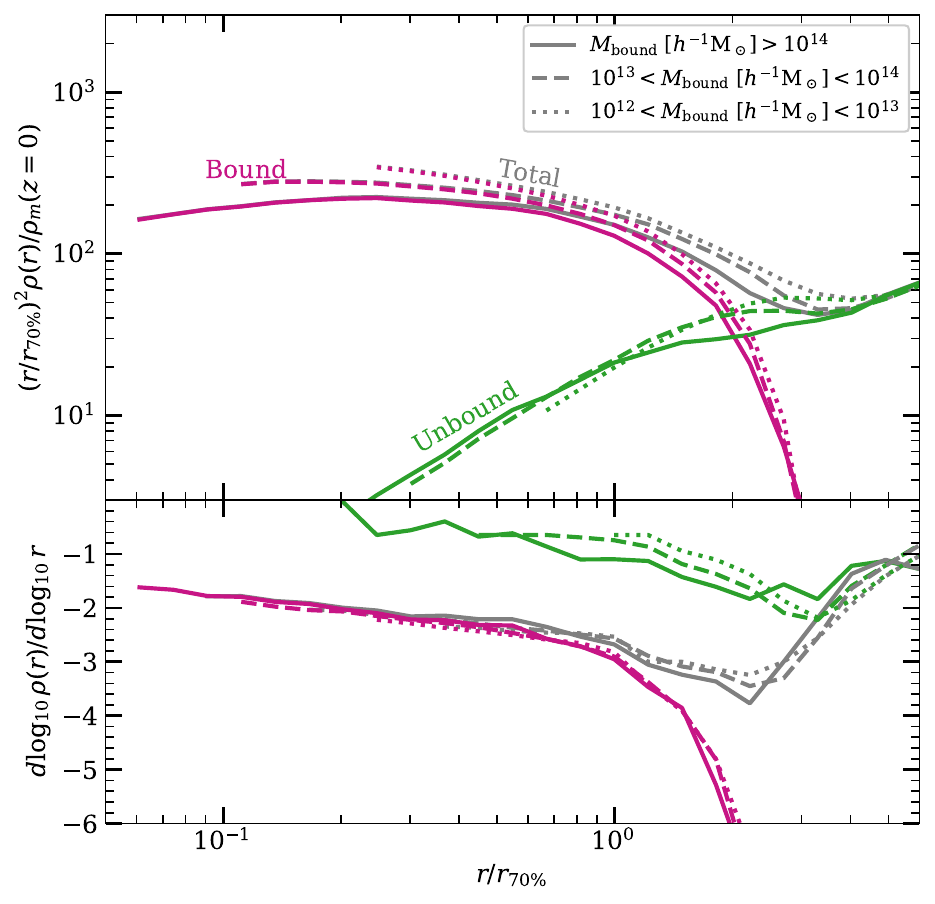}
    \caption{Upper panel: Median matter density profiles of haloes split into bound (pink), unbound (green), and all (grey) particles using {\sc strawberry} for all haloes in three mass bins. Lower panel: Median logarithmic derivative of the matter density profile split into bound (pink), unbound (green), and all (grey) particles using {\sc strawberry} for three mass bins. The axes are normalised with respect to the radius, $r_{70\%}$, containing 70\% of the bound mass. The bound and total profiles are cut at six times the softening length of the simulation. The unbound profiles are only shown for bins containing more than ten particles.}
    \label{fig:dens_profiles}
\end{figure}

In the upper panel of Fig.\ref{fig:dens_profiles}, we show the average matter density profile of bound and unbound particles, respectively, as pink and green lines. Here, the unbound population is defined as all particles around the halo that are not considered bound, meaning that we also include particles `outside' the potential well. Each line style corresponds to the median profile over different mass ranges, as indicated by the legend. 
Both axes are normalised with respect to the radius, $r_{70\%}$, containing 70\% of the bound mass.  

We see in Fig.~\ref{fig:dens_profiles}, that all bound profiles exhibit a similar behaviour. Indeed, for all mass bins, we observe that up to roughly $r_{70\%}$, the slope of the profile varies slowly as a function of radius. Beyond this point, the profile steepens rapidly, exhibiting an exponential cut-off. As such, we find that haloes possess a clear-cut edge, beyond which particles are not bound to the structure. This boundary results in a convergent halo mass, justifying the definition of a bound mass. This boundary is, however, not directly visible in the density profile and can only be obtained in terms of particle dynamics. Indeed, this result is analogous to that of \citet{Diemer2022}, where the author obtains an exponential cut off in the profile when selecting only particles that have passed their first pericentre. In combination with Fig.~\ref{fig:binding_prob}, this is unsurprising, as the fraction of bound particles rises sharply as it approaches its first pericentric passage, meaning that in practice both approaches result in a similar selection of particles.

To further quantify the evolution of the slope of the density profile, in the lower panel of Fig.~\ref{fig:dens_profiles}, we show the median logarithmic slope, $\gamma = \dd \log_{10}\rho(r)/\dd\log_{10}r$, for each mass bin. For the bound population, the slope decays slowly from $\gamma = -1$ at $r=0$ to $\gamma\simeq-3$ at $r= r_{70\%}$, which is consistent with the NFW or Einasto profile. Further from the centre, on the other hand, we observe an exponential decay of the profile, with the logarithmic slope plunging rapidly. Beyond this point, the grey curves representing the logarithmic slope of the total profile, i.e. including both bound and unbound particles, detach from the bound population. We observe that the density of the unbound population decreases at a slower rate than its bound counterpart, becoming dominant farther from the centre, with the total profile beginning to significantly depart from the bound population around $2\,r_{70\%}$. Beyond the the exponential cut-off in the bound profile, the logarithmic slope of the total population remains above $\gamma \gtrsim -3$, before presenting a dip, also known as the splashback feature \citep{Adhikari2014}, between 2 and 3 times $r_{70\%}$ and increasing again.

\begin{figure}
    \centering
    \includegraphics[width=0.95\linewidth]{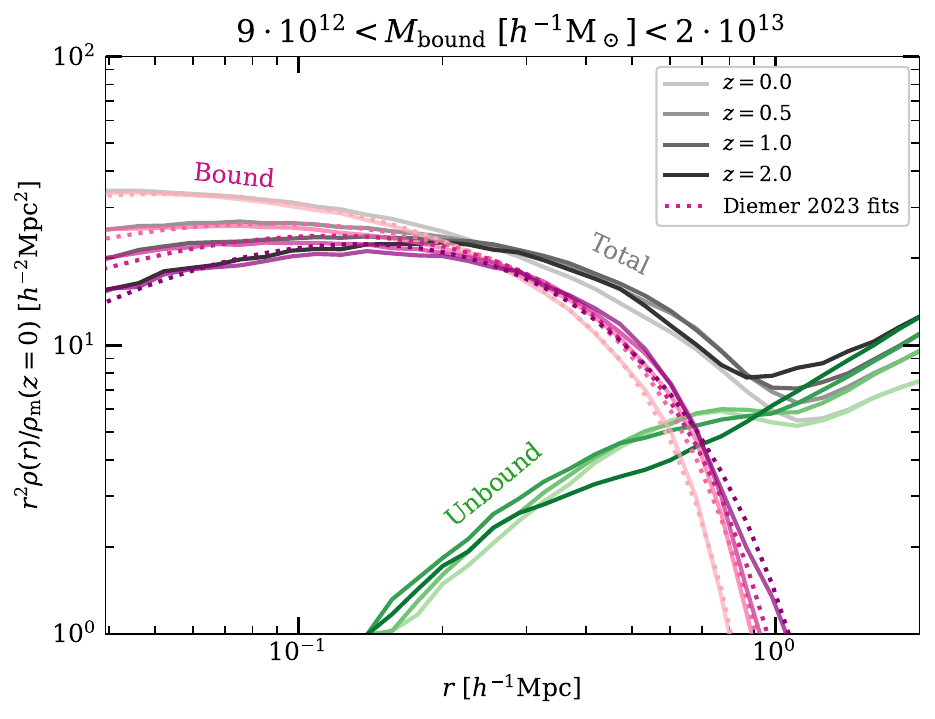}
    \caption{Median comoving matter density profile of haloes with bound masses, $9\cdot10^{12}\hMsun$ and $3\cdot10^{13}\hMsun$. For each redshift, we separate the total profile, in grey, into three components: the bound population, in magenta, the unbound population, in green, and particles outside the potential well, in blue. For each line colour, higher redshifts are marked by darker colours. The dotted pink lines represent the best fit \citet{Diemer2023} model B profiles to the bound population.} 
    \label{fig:profile_redshift}
\end{figure}

In Fig.~\ref{fig:profile_redshift} we plot the same decomposition of the halo profile, but for a single tighter mass range, $9\cdot10^{12}\hMsun< M_{\rm bound} < 2\cdot10^{13}\hMsun$, and multiple redshifts, $z=0,\ 0.5,\ 1,$ and $z=2$. Interestingly, we observe that the individual components of the profile evolve differently. We find that the outskirts of the bound population change very little with time. This is despite an apparent change to the inner slope and hints at a tight correlation between the bound mass and size of the bound region. Conversely, we find the evolution of the total profile is substantial in this region and is primarily dominated by the evolution of the unbound population. We observe that as the background density decreases with decreasing redshift, so does the density of infalling material surrounding the halo.

In this same figure, we fit the \citet{Diemer2023} model B profile,
\begin{align}
    \rho(r) &= \rho_0\exp[S(r,r_{\rm s},r_{\rm t},\alpha,\beta)],\nonumber\\
    S &= - \frac{2}{\alpha}\left[\left(\frac{r}{r_{\rm s}}\right)^\alpha - 1\right] - \frac{1}{\beta}\left[\left(\frac{r}{r_{\rm t}}\right)^\beta - \left(\frac{r_{\rm s}}{r_{\rm t}}\right)^\beta\right]\\\
    &+ \frac{1}{\eta}\left(\frac{r_{\rm s}}{r_{\rm t}}\right)^\beta \left[\left(\frac{r}{r_{\rm s}}\right)^\eta + 1\right], \nonumber
    \label{eq:Diemer23_modB}
\end{align}
finding a good agreement between the parametric form and the density profile of the bound populations. Moreover, we find that the resulting best fit parameters are consistent with the ranges quoted in \citet{Diemer2025}, indicating that it is possible to distinguish between the orbiting and infalling populations, as defined in these works, using purely instantaneous information, meaning without the need to reconstruct particle orbits. For completeness, in App.~\ref{app:best_fit} we fit this model to the density profiles of individual haloes and present the resulting best fit parameters.

\subsection{Virialisation}
\label{sec:virial}

In theoretical settings, haloes are considered to be stable, gravitationally bound structures. To have a system that remains finite over long time spans it is required that it fulfils the virial theorem:
\begin{equation}
   2T + G = \langle{\bf v}\cdot{\bf v}\rangle +\langle{\bf r}\cdot{\bf a}\rangle = 0,
\end{equation}
where $G$ is the virial, $T$ is the mean kinetic energy, and ${\bf r}$, ${\bf v}$, and {$\bf a$} are all defined as having vanishing means. Here we use the more fundamental `force' virial theorem, as it does not require any additional assumptions, which are needed to define the potential energy used in the more common `potential' virial theorem.
In its original form, the brackets represent time averages performed over the trajectory of single particles. However, under the assumption that the population is fixed and ergodic, these time averages can be replaced by averages over a sufficiently large number of $N$-body particles sampling different orbits inside the system. 

In simulations, it has been shown that particle distributions recovered by classical halo finding algorithms do not fulfil the virial condition, typically underpredicting the ratio $-G/T$ \citep[e.g.][]{Shaw2006, Poole2006, Davis2011}. This effect is understood to be caused by the unintended inclusion of unbound particles in the outskirts of haloes \citep{Shaw2006}. Indeed, as discussed in Sect.~\ref{sec:halo_props}, these particles have not yet mixed with the bound population and tend to have larger kinetic energies while being located at larger radii. As such their inclusion increases the average kinetic energy, $T$, and thus decreases the virial ratio. 

\begin{figure}
    \centering
    \includegraphics[width=0.95\linewidth]{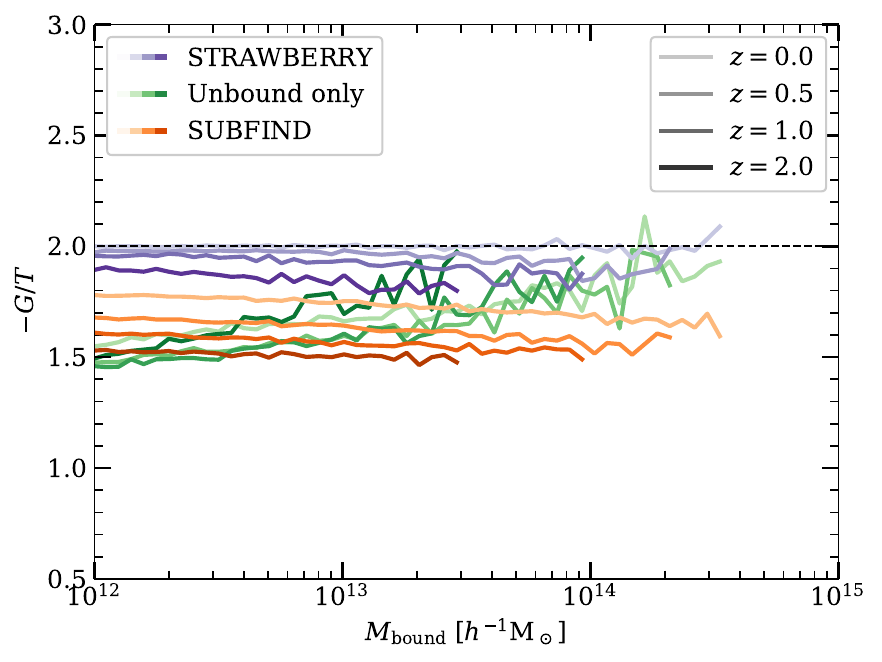}
    \caption{Median viral ratios as a function of mass, in purple for {\sc strawberry} selected particle distributions, in green particles unbound by {\sc strawberry}, and in orange for {\sc subfind} selected particle distributions. The brightness of each curve corresponds to the redshift of the sample, with darker tones representing higher redshifts.}
    \label{fig:G_T_distrib}
\end{figure}

The result of this effect can be seen in Fig.~\ref{fig:G_T_distrib} where the orange curves represent the median virial ratio measured from {\sc subfind} particle distributions. 
Here the aforementioned offset is clearly visible, with the median virial ratio being approximately 10\% to 25\% below the expected value, depending on the redshift. We note that the redshift evolution of these curves is in line with our previous observations that haloes on average tend to be surrounded by more infalling material at higher redshifts, thus resulting in a larger bias towards high kinetic energies.

In \citet{Stucker2021}, the authors show that the bound population at $z=0$ is virialised with the resulting virial ratio being in agreement with the theoretical prediction. Here, we expand on these previous results and also show the redshift evolution.  
Indeed, the {\sc strawberry} selected particles, shown in purple, yield a median in agreement with the theoretical expectation. We note a residual shift towards lower values at high masses and high redshifts which we attribute to recently formed, accretion dominated haloes which have not had time to fully relax. For the purpose of visual clarity, we have not shown the dispersion around the median, which is roughly constant of the order of 10\%. 
Furthermore, if we plot the median virial ratios estimated only using the particles considered unbound by {\sc strawberry} (green lines in Fig.~\ref{fig:G_T_distrib}) we observe a very strong bias toward lower values, and also recover the redshift trend seen in the {\sc subfind} selected particles. This confirms that it is indeed the inclusion of unbound particles which produces the decrease seen in classical halo finders. 

\subsection{Bound masses}
\label{sec:pop_props}

\begin{figure}
    \centering
    \includegraphics[width=0.95\linewidth]{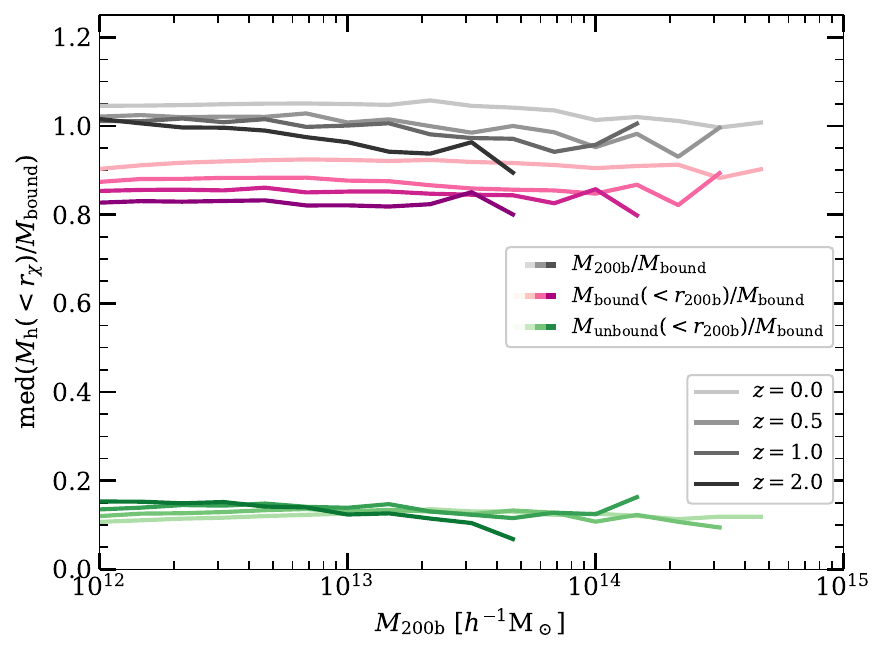}
    \caption{Median mass ratios as a function of $M_{\rm 200b}$. The colour of the lines correspond to the ratio en question, in grey $M_{\rm 200b}/M_{\rm bound}$, and in pink $M_{\rm bound}(<r_{\rm 200b})/M_{\rm bound}$. The shade of each line corresponds to the redshift, specifically, from brightest to darkest, $z=0,\ 0.5,\ 1.0,$ and $2.0$.}
    \label{fig:mass_ratio_redshift}
\end{figure}

In this final section, we investigate how our previously introduced mass definition, $M_{\rm bound}$, relates to the wider ecosystem of masses used in cosmology. Indeed, \citet{Stucker2021}, observe that, at $z=0$, the bound mass appears to closely track the SO mass $M_{\rm 200b}$. Here, we wish to see if this is also the case at higher redshift. To begin, we estimate the median mass ratio $M_{\rm 200b}/M_{\rm bound}$ as a function of $M_{\rm 200b}$.  This ratio, shown as grey lines in Fig.~\ref{fig:mass_ratio_redshift}, is computed for four redshifts, $z=0,\ 0.5,\ 1.0,$ and $2.0$, corresponding with the shade of each line, respectively from brightest to darkest. We plot the resulting curves omitting the scatter around these measurements for visual clarity, the latter being roughly constant, $\sigma\simeq0.2$, for all masses and redshift.

At redshift $z=0$, we recover that the bound mass is comparable to $M_{\rm 200 b}$. At higher redshifts, we see that while both masses remain qualitatively similar, the ratio does shift to a lower value. This trend is driven by the bound region being more extended at higher redshifts, which can also be seen in Fig.~\ref{fig:profile_redshift}. We can further observe this trend by plotting the ratio of the amount of bound mass within $r_{\rm 200 b}$ with respect to the total bound mass, in pink, using the same brightness convention as previously. In doing so, we can clearly observe that this ratio increases with redshift. This indicates that at higher redshifts $r_{\rm 200 b}$ contains less bound material than at lower redshift, reinforcing the observation that the bound region is more extended at higher redshifts.

To further understand the origin of this trend, we plot the median fraction of unbound material within $r_{200\rm b}$, in green, following the same convention for higher redshifts. In this case, the redshift dependence appears to be less visible. This indicates that while $M_{\rm 200 b}$ will primarily contain bound material, the physical region it will be probing will change as a function of redshift. A possible explanation for the observed trend is that $r_{\rm 200 b}$ is defined with respect to the total matter density and cosmological background. As such, it will not only depend on how matter is distributed inside the halo but also on the cosmological background evolution. This creates a compounding effect where changes in the density profile, as seen in Fig.~\ref{fig:profile_redshift}, and cosmological background both change the definition of $r_{200\rm b}$ \citep[see e.g.][]{Diemer2014}.

\section{Discussion}
\label{sec:discussion}

In this work, we have presented a novel algorithm for the assignment of particles to haloes in cosmological simulations. This algorithm, operating on the boosted gravitational potential \citep{Stucker2021}, defines a halo as a set of particles that is energetically confined to a potential well.
In this context, gravitational binding is determined by directly comparing the energy of particles to the escape energy of the halo given by the saddle point in the potential that leads to a lower minimum. As such, this can be seen as a self consistent and parameter free definition of haloes.
Particles with lower energies are kept and considered bound to the structure, while particles with higher energies are considered unbound as they would theoretically be able to exit through the saddle point. This algorithm presents a considerable improvement with respect to that used by \cite{Stucker2021}, being both more numerically efficient and taking into account additional effects, for instance in the handling of energies on large scales and at redshifts $z\neq0$, allowing for a more in depth analysis.

Equipped with a clear binding-notion, we have studied how particles become bound to haloes. In this formalism, we find that despite the fact that the energy of particles is continuously increasing, particles become bound because the saddle point energy increases faster. Moreover, we find that once a particle is accreted, binding takes place on a short time scale, typically only in one dynamical time . This observation echoes the particle selection criteria of \cite{Adhikari2014} and \cite{Sparta2017}, where orbiting particles are selected based on whether they have passed a given point in their first orbit.

Similarly to previous studies \citep[e.g.][]{Wang2011}, we observe that newly accreted particles are deposited on large orbits in the outskirts of the halo. In addition, we find that the bound population is virialised and exhibits a well defined edge. On the other hand, the unbound population, which inhabits the outskirts, is not virialised and does not exhibit such an edge and even obscures the edge of the bound population when looking at the density profile of all particles surrounding the halo. In phase space and energy space, we find that this edge separates the bound population form the infalling and first splashback streams, qualitatively corresponding to a feature previously used to define the orbiting population \citep{Garcia2023,Salazar2024}.
 
At higher redshifts, as seen in Fig.~\ref{fig:profile_redshift}, we identify distinct evolution mechanisms for both populations. At similar bound masses, the bound component does not exhibit significant differences at the edge, with the extent of the bound region remaining relatively stable. In contrast, the evolution of the total profile is substantial in this region, being mainly driven by the unbound population. As such, in the outskirts the evolution is driven by external factors such as the accretion rate, the background density and external tidal fields.

This work paves the way for deeper studies of dark matter haloes. In particular, in cases where a clear definition in terms of energies is advantageous. For instance, when creating idealised initial conditions for haloes, where  the relation between the density and gravitational potential is used to perform the Eddington inversion and obtain the particle distribution function \citep[e.g.][]{Binney2008}.
Alternatively, the definition of haloes outlined here can be of interest when a distinction between a virialised and non-virialised state is needed. This can be the case when studying how gas settles in the potential well, or how tidal stripping affects haloes. Finally, \citet{Ondaro2022} associate deviations from universality in the halo mass function defined at $M_{\rm 200 b}$ to the mass accretion history of haloes, given the stability of the bound profile seen in Sect.~\ref{sec:halo_profile}, it would be of interest to investigate further whether $M_{\rm bound}$ provides a more universal alternative to standard spherical overdensity masses.

\section*{Data availability}

The codes designed for and used throughout this work along with additional examples are publicly available on GitHub: \href{https://github.com/trg-richardson/strawberry}{https://github.com/trg-richardson/strawberry}.

\begin{acknowledgements}

The authors would like to thank Eduardo Rozo, Edgar Salazar, Lurdes Ondaro-Mallea, and the anonymous referee for their insightful comments. This project has received funding from the Spanish Government's grant program ``Proyectos de Generaci\'on de Conocimiento'' under grant number PID2021-128338NB-I00. JS acknowledges funding by the Austrian Science Fund (FWF) [10.55776/ESP705]. Many of the numerical computations carried out, and plots presented in this work were made possible by the SciPy \citep{scipy}, NumPy \citep{numpy}, Matplotlib \citep{matplotlib}, and Bacco \citep{bacco} packages.

\end{acknowledgements}

\bibliographystyle{aa}
\bibliography{bibliography}

\appendix

\section{Convergence}
\label{app:scaling_convergence}

This appendix is devoted to validating the various numerical aspects of the algorithm presented in Sect.~\ref{sec:algorithm}, and provide additional specifics regarding the implementation of the latter. Most notably, we first test the effect of numerical resolution on the quantities presented throughout this work before investigating the effect of the two main design choices, namely the number of neighbours assigned to each particle and the impact of the large scale potential dampening factor, $\tilde\delta$. 

\subsection{Simulation resolution}

We first test the effect of resolution.
To visualise this
we measure the HMF and median persistence
at redshift, $z=0$, as estimated in three simulations which differ solely by the number of particles they contain. To do so, in addition to the simulation used throughout the main text we run two lower resolution simulations, respectively with $N = 156^3$, and $ N = 312^3$ particles, using the same realisation of the Gaussian random noise field.

\begin{figure}
    \centering
    \includegraphics[width=0.95\linewidth]{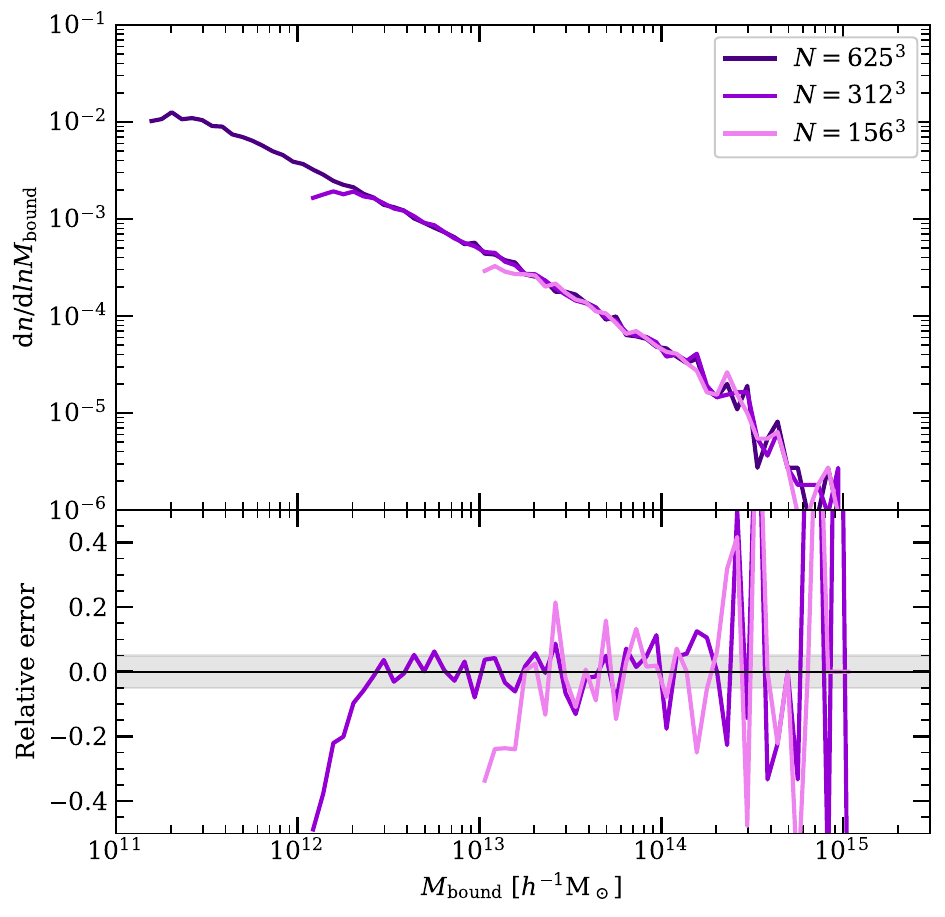}
    \caption{Upper panel: bound halo mass function as estimated in three simulations with the same cosmology but differing particle numbers. Lower panel: relative error with respect to the highest resolution ($N=625^3$) simulation. The shaded region corresponds a 5\% deviation.}
    \label{fig:hmf_resolution}
\end{figure}

In Fig.~\ref{fig:hmf_resolution}, we plot the HMF estimated from the measured bound masses.
We observe that the HMF converges for haloes containing more than a few hundred particles. Below this mass, we see a significant difference between the higher and lower resolution simulations. This is due to two effects, one intrinsic to the resolution and a second selection effect. The first effect is caused by the loss of spatial resolution in the inner regions of haloes. This reduces the finesse with which the algorithm can locate the minimum in the potential, effectively smoothing out the inner cusp. This smoothing, results in a lower persistence and as such a lower mass. 
The second effect mentioned above comes from the choice of which seeds are or are not considered starting points for the algorithm. Indeed, here, we only consider haloes detected using {\sc subfind} as potential seeds if they contain more than 50 particles. These haloes will have a range of bound masses, meaning that the resulting selection in terms of bound halo masses introduces a systematic bias at the low mass end, and is in fact dominant here. 

\begin{figure}
    \centering
    \includegraphics[width=0.95\linewidth]{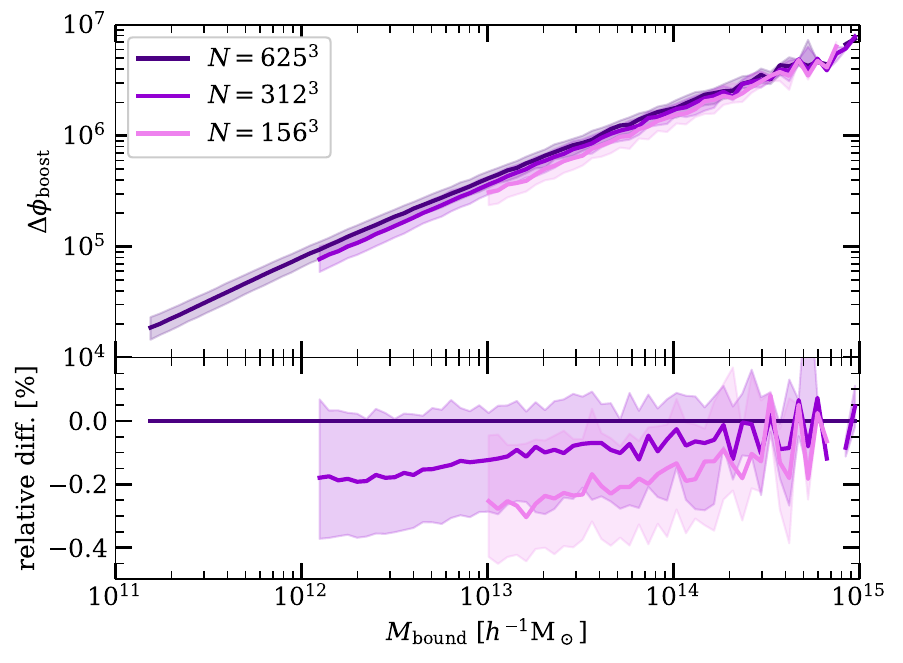}
    \caption{Upper panel: Median persistence as a function of halo mass for three particle resolutions. For each curve, the shaded region contains 68\% of haloes in the corresponding mass bin.Lower panel: relative difference with respect to the median persistence in the highest resolution simulation.}
    \label{fig:persistance_resolution}
\end{figure}

The first effects becomes more apparent when plotting the median persistence as a function mass as shown in Fig.~\ref{fig:persistance_resolution}.
Indeed, as stated above, this is because the central cusp of the potential is sampled with fewer particles.
This results in an underestimation of the potential depth when using lower resolution simulations, but also for smaller haloes which are sampled by fewer particles. In Fig.~\ref{fig:persistance_resolution}, this is visualised by the slow convergence of the curves at high masses and high resolution. While this does not significantly affect the recovered masses, as evidenced by the absence of bias in Fig.~\ref{fig:hmf_resolution}, it is clear that resolution effects should be taken into account when analysing these persistences, with an observed relative difference of the order 20\% between the median persistence of estimated using the highest resolution simulation and both lower simulations.

\subsection{Number of neighbours}
\label{app:neighbours}

\begin{figure}
    \centering
    \includegraphics[width=0.95\linewidth]{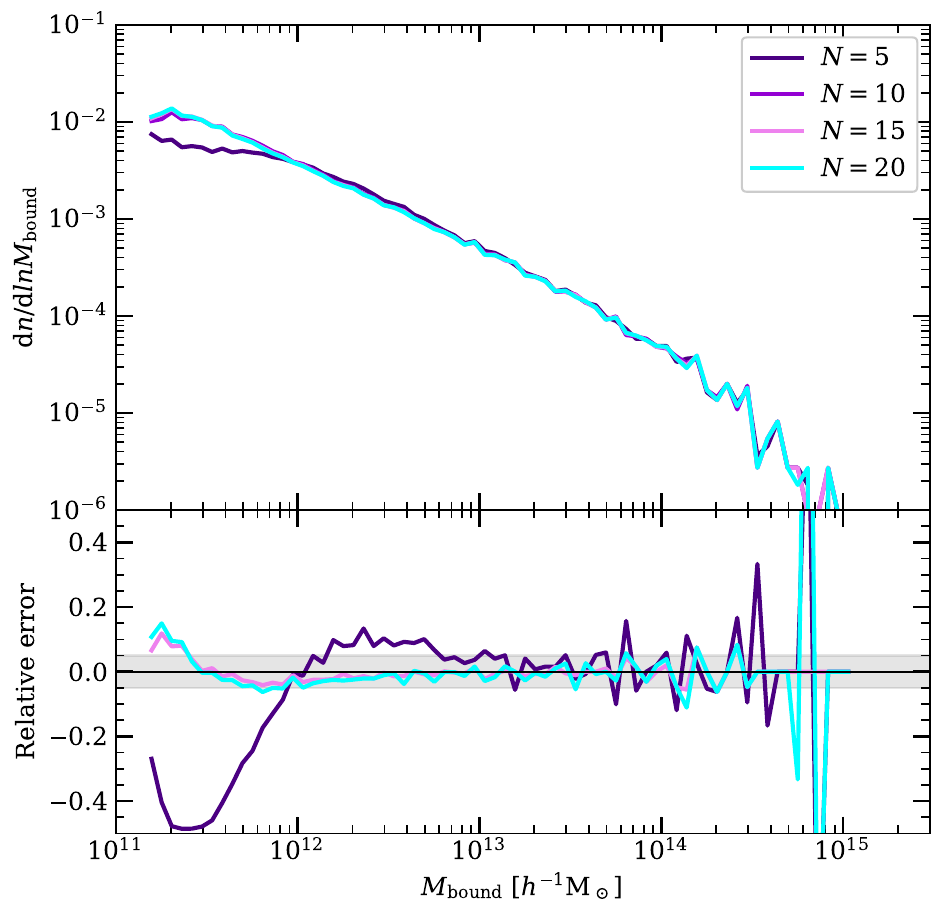}
    \caption{Upper panel: bound halo mass function as estimated using {\sc strawberry} outputs on a single simulation but changing the number of nearest neighbours used by the algorithm. Lower panel: relative error with respect to the nominal value ($N=10$) of nearest neighbours. The shaded region corresponds a 5\% deviation.}
    \label{fig:hmf_neighbours}
\end{figure}

Let us now examine the algorithmic choices that are made in Sect.~\ref{sec:algorithm}, starting with the number of neighbours we assign to each particle. To do so, we repeat the same exercise as above but changing the number of neighbours assigned to each particle. In the context of this work, we have chosen to limit ourselves to considering a fixed number of nearest neighbours, $N=10$, for all particles. 
To justify this decision, in Fig.~\ref{fig:hmf_neighbours} we estimate the HMF for four different numbers of nearest neighbours, specifically $N=5,\ 10,\ 15,$ and $20$. At high masses, we observe a general agreement between all the curves. However, at low masses we observe a significant divergence of the $N=5$ HMF, 
with this particular choice of neighbours under-predicting the abundance of haloes in this masse range. The main systematic in this case, is a clumping effect, which hinders exploration of the medium in low density regions. 
We can see this in Fig.~\ref{fig:hmf_neighbours} where there is good convergence passed $N=10$ but for $N=5$ we overestimate the abundance of intermediate mass haloes and underestimate the abundance of the low mass counterparts.  
Beyond $N=10$, we do not observe any significant changes that would justify the additional computational resources needed for these connectivity definitions.
Given these observations, we decide to use the $N=10$ nearest neighbours of each particle to define the connectivity.

\subsection{Large scale potential dampening}
\label{app:dep_delta}

As discussed in the main-text, the definition of the gravitational potential in an expanding (and uniformly accelerated) universe inevitably leaves a degree of freedom associated to what should be considered `at rest'. While we have proposed the pragmatic choice of $\tilde\delta = 4.55$, targetting the turn-around radius in spherical collapse scenarios in EdS universes as a saddle-point, we show here that the precise value of this has a very minor impact on the haloes that are identified.

\begin{figure}
    \centering
    \includegraphics[width=0.95\linewidth]{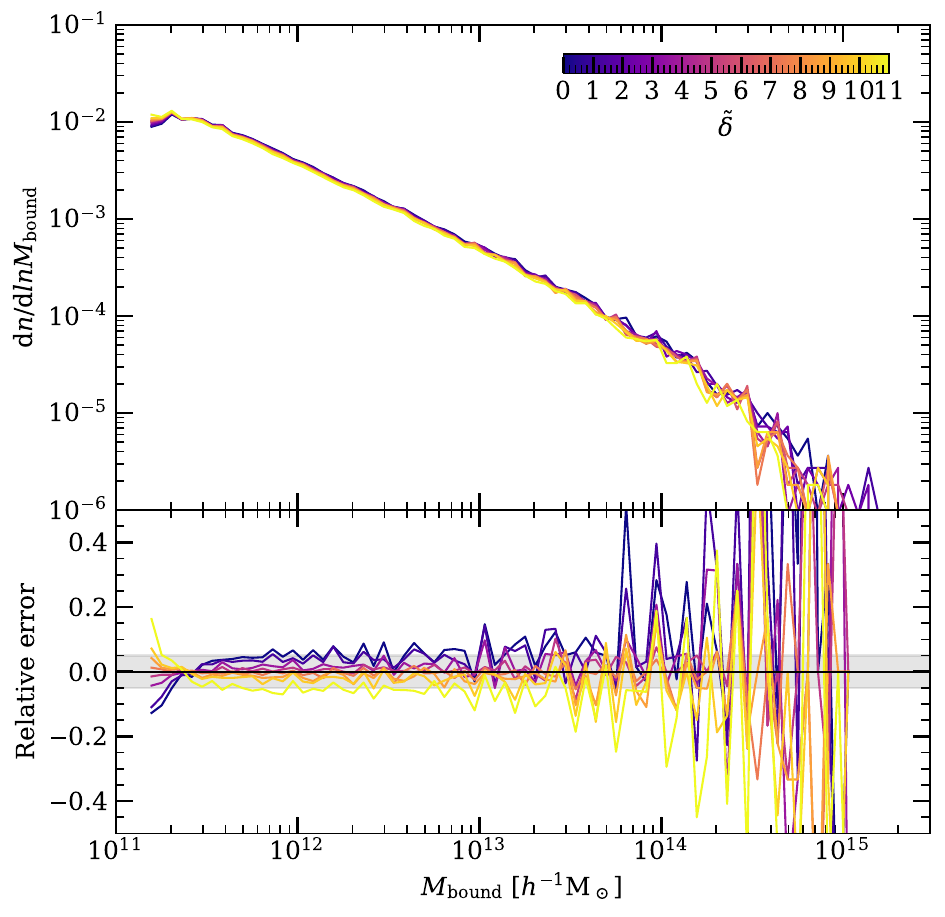}
    \caption{Upper panel: bound halo mass function as estimated using {\sc strawberry} outputs on a single simulation using ten different values of the large scale parameter, $\tilde\delta$, ranging from 0 to 10.939. Lower panel: relative difference with respect to halo mass function estimated using the nominal value ($\tilde\delta = 4.55165$) used in this work. The shaded region corresponds a 5\% deviation.}
    \label{fig:hmf_dep_delta}
\end{figure}

It is however important to note that this decision does have an influence on the estimated physical quantities and as such is worth analysing in further detail. To do so, we repeat the measurements presented above for ten values of $\tilde\delta$ ranging from 0 to 10.939\footnote{This specific value corresponding to the turn-around density contrast at $z=0$, for the cosmology of the simulation.}. Starting with the HMF shown in Fig.~\ref{fig:hmf_dep_delta}. Here, we observe that depending on the value of the parameter, the abundance of haloes of a given mass can be either underestimated or overestimated up to the order of 5\% to 10\%. This is indicative that while the large scale correction is designed to operate only on the largest haloes, it impacts haloes at all masses. This is because, for small and intermediate mass haloes the correction still lowers the level of the saddle point, even if its contribution is subdominant.
As a result, certain particles are considered unbound because their mechanical energy falls between the physical potential depth and the artificially lowered value. In turn, this shifts the HMF towards lower masses, as is observed here.

\begin{figure}
    \centering
    \includegraphics[width=0.95\linewidth]{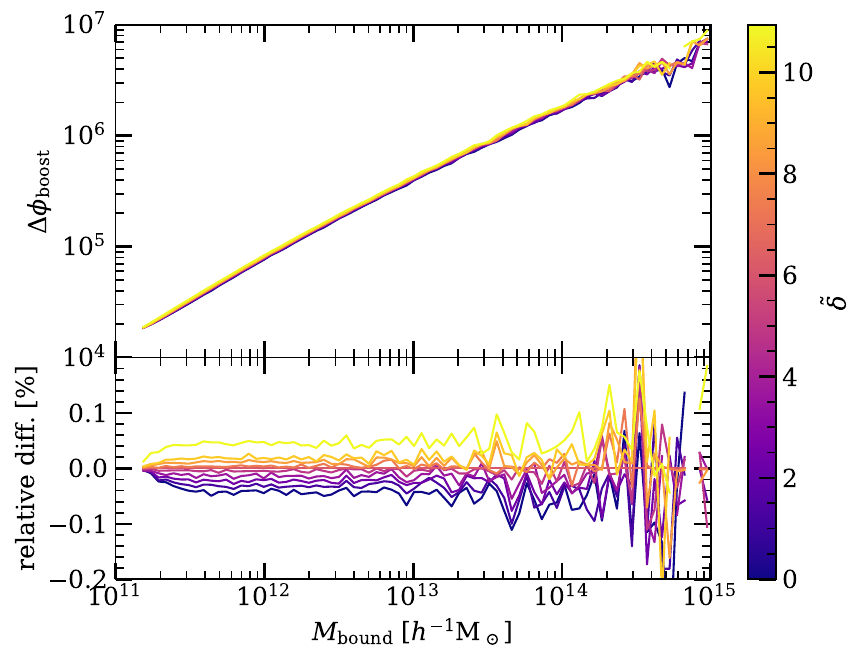}
    \caption{Upper panel: Median persistence as a function of bound halo mass for ten values of $\tilde\delta$ ranging from 0 to 10.939. Lower panel: relative difference with respect to the median persistence estimated for $\tilde\delta=4.55$.}
    \label{fig:persistance_dep_delta}
\end{figure}

Regarding the effect $\tilde\delta$ has on the persistence, we observe in Fig.~\ref{fig:persistance_dep_delta} a 5\% shift of the $\Delta\phi_{b,*} - M_{\rm bound}$ relation. As the persistence is corrected for $\tilde\delta$, this shift is indicative of higher persistence haloes being assigned lower masses for larger values of large scale parameter. While this deviation is smaller than the impact of the resolution on the persistence it is nonetheless important to keep in mind when studying the persistences.

\section{Profile fitting parameters}
\label{app:best_fit}

\begin{figure}
    \centering
    \includegraphics[width=0.95\linewidth]{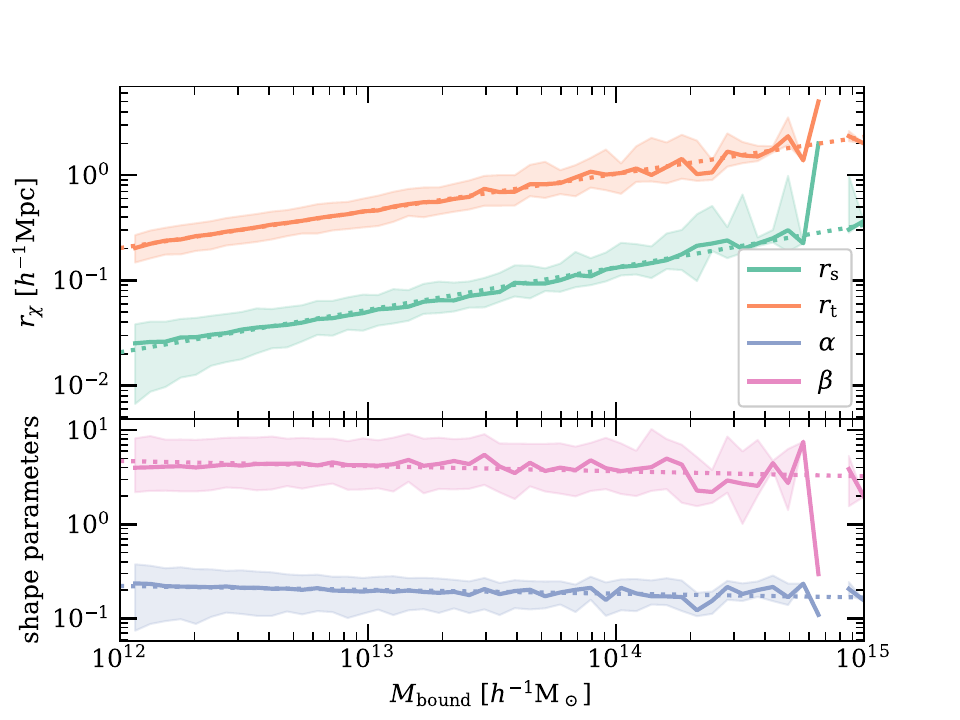}
    \caption{Median and 1-$\sigma$ dispersion of best fit  \citet{Diemer2023} model B profile parameters as a function of bound mass, respectively solid lines and shaded regions. Dotted lines represent the best fit power law to the median.}
    \label{fig:diemer_fits}
\end{figure}

As stated in the main text we find that, on average, the \citet{Diemer2023} model B profile provides a good fit to the density profile of the bound population. In this appendix, we provide an expanded view where we fit this model to the individual log profiles  of haloes using a standard $\chi^2$ minimisation procedure. 

In Fig.~\ref{fig:diemer_fits} we plot the distributions of resulting fitting parameters as a function of the bound mass. We observe that both the scale radius and truncation radius vary significantly with bound mass. We quantify this trend by fitting a power-law to the median fitted values, excluding the largest mass bins which are dominated by only a few high mass haloes. The best fit power laws,
\begin{align}
    r_{\rm s} &= (0.05 \pm 0.01)\left[M_{\rm bound}/10^{13}\hMsun\right]^{0.403 \pm 0.007}\,h^{-1}{\rm Mpc},\\
 r_{\rm t} &= (0.46 \pm 0.01)\left[M_{\rm bound}/10^{13}\hMsun\right]^{0.352 \pm 0.007}\,h^{-1}{\rm Mpc},
\end{align}

are shown as dotted lines in the upper panel of Fig.~\ref{fig:diemer_fits}. In contrast, and as also observed in \citet{Diemer2025}, we find that the two shape parameters,
\begin{align}
    \alpha & = (0.20 \pm 0.01)\left[M_{\rm bound}/10^{13}\hMsun\right]^{-0.041 \pm 0.008}\;\text{ and} \\
    \beta &= (4.15 \pm 0.03)\left[M_{\rm bound}/10^{13}\hMsun\right]^{-0.05\pm 0.01},
\end{align}
vary only slowly as a function of the bound mass.

A full quantification of the behaviour of these parameters as a function of redshift and cosmology, is out of the scope of this work and requires a separate dedicated study. Nonetheless, the ranges of parameters is compatible with those found in \citet{Diemer2025}, which can be attributed to the similarity between the \textsc{sparta} orbital selection and our equivalent effective selection as illustrated in Fig.~\ref{fig:binding_prob}.

\end{document}